\documentclass[twocolumn]{aastex631}  
\usepackage{enumitem}
\usepackage{amsmath}
\usepackage{cases}
\usepackage{xcolor}
\usepackage{tablefootnote}
\usepackage{subfigure}
\usepackage{bm}
\hypersetup{backref=true,pagebackref=true,  hyperindex=true,colorlinks=blue,breaklinks=true,  urlcolor=violet,linkcolor= red, bookmarks=true, 	bookmarksopen=true,  filecolor=cyan, citecolor=blue, linkbordercolor=blue}
\shorttitle{}
\shortauthors{}

\begin{document}
\title{Simulating Gravitational Microlensing Events by TESS: Predictions on Statistics and Properties}

\author[0000-0002-0167-3595]{Sedighe Sajadian}
\affiliation{Department of Physics, Isfahan University of Technology, Isfahan 84156-83111, Iran; \url{s.sajadian@iut.ac.ir}}

\author[0009-0006-2285-6792]{Atousa Kalantari}
\affiliation{Department of Physics, Institute for advanced Studies in Basic Science (IASBS), Zanjan 45137-66731, Iran}

\author[0000-0002-7611-9249]{Hossein Fatheddin} 
\affiliation{Leiden Observatory, Leiden University, PO Box 9513, NL-2300 RA Leiden, the Netherlands}

\author[0000-0002-1910-7065]{Somayeh Khakpash}
\affiliation{Rutgers University, Department of Physics \& Astronomy, 136 Frelinghuysen Rd, Piscataway, NJ 08854, USA}

\begin{abstract}
We study the statistics and properties of microlensing events that can be detected by the Transiting Exoplanet Survey Satellite(TESS) based on Monte Carlo simulations. We simulate potential microlensing events from a sample of the TESS Candidate Target List(CTL) stars by assuming different observational time spans(or different numbers of sectors for each star) and a wide range of lens masses, i.e., $M_{\rm l}\in [0.1M_{\oplus},~2 M_{\odot}]$. On average, the microlensing optical depth and the event rate for CTL stars are $\simeq 0.2\times 10^{-9}$, and $\Gamma_{\rm{TESS}}\simeq0.6\times10^{-9}$ per star per day, respectively. The microlensing optical depth decreases by increasing the CTL priority, whereas the efficiency for detecting their microlensing signals enhances with the priority. Additionally, we simulate the microlensing events from the TESS Full-Frame Images(FFIs) stars extracted from the \texttt{TESS}-\texttt{SPOC} pipeline. The optical depth and event rate for these stars are on average $\simeq 1$-$3\times 10^{-9}$, and $\Gamma_{\rm{TESS}}\simeq 1$-$4\times 10^{-9}$ per star per day, and their highest values occur for sector $12$. The total number of microlensing events for the CTL stars is $N_{\rm e, \rm{tot}}\sim0.03$, whereas for the FFIs' stars number of events per star during $27.4$-day observing windows is $\hat{N}_{\rm e, \rm{tot}}\simeq1.4 \times 10^{-6}$. Based on four criteria we extract the detectable microlensing events and evaluate the detection efficiencies. The highest efficiency for detecting microlensing events from the TESS data occurs for the lens mass $\log_{10}[M_{\rm l}(M_{\odot})] \in [-4.5$,~$-2.5]$, i.e., super-Earth to Jupiter-mass Free Floating Planets(FFPs). The detectable microlensing events from the TESS stars are significantly affected by both finite-source and parallax effects. 
\end{abstract}


\section{Introduction}
Transiting Exoplanet Survey Satellite (TESS)\footnote{\url{https://tess.mit.edu/}} is a telescope designed by the NASA Explorer mission which was launched on $18$ April $2018$. This space telescope is orbiting the Earth in a highly eccentric orbit with the period of $13.7$ days. Hence, it is rotating the Earth twice during one lunar period. Every $13.7$ days while this telescope is in its orbital perigee, it transfers its collected data to the Earth \citep[][]{2014tesspaper}. The TESS telescope was planned to scan each hemisphere during one year. Each hemisphere is divided into $13$ sectors. Each sector extends $24\times 96$ square degrees covered by four wide-field cameras with a field of view $24\times 24$ square degrees, so that they are overlapped at the ecliptic poles. Every sector is observed during two $13.7$-day orbits \citep{2015TESSstart}.

The main goal of the TESS mission was detecting small and near extra solar planets through dense and highly accurate photometric measurements of close and bright stars. The TESS telescope has observed (and observes) around $200$k-$400$k stars, the so-called Candidate Target List (CTL), selected from brightest and closest stars with pre-measured physical parameters planned to be observed with a two-minute cadence \citep{2018AJstassun,2019AJstassun}. Additionally, its full frame images (FFIs), produced from co-adding $2$-second images over either $10$ or $30$ or $3.3$ minutes, are released publicly. By applying the MIT Quick Look Pipeline (QLP) on these FFIs, \citet{2020HuangFFI} presented the light curves of $15$ and $10$ million stars from the Southern and Northern hemispheres, respectively. We note that the ecliptic poles (overlapped parts of all sectors) are observed during around one year continuously. These huge photometric data from bright stars will offer a unique opportunity to search for either transient or periodic astrophysical signals with short time scales (around one month or less) toward different lines of sight.  

\citet{2018ApJSBarclay} predicted that this telescope would discover $\sim 1250$ and $3100$ transiting exo-planets from observing CTL stars during two years with a two-minute cadence, and FFIs stars, respectively. Other estimations for the number of planets expected to be found by TESS are within the same ranges \citep{2015ApJSullivan,2018Huang}. Additionally, \citet{2019AJVillanueva} estimated that the number of single-transit planetary events would be $241$ and $977$ from two-minute observations and the light curves from FFIs, respectively. The important advantage of the transit method for discovering extra solar planets is plausible follow-up spectroscopic observations (even ground-based ones) to infer the planet physical parameters. Up to now more than $4216$ extra solar planets have been discovered by this method whereas all discovered planets are $5678$ \citep[As of 2024-07-02,][]{mastexoplanet}. Nevertheless, this method can only detect bound planets in close orbits with edge-on orbital planes as seen by the observer.    
  
  
Detecting unbound and dark planets, the so-called Free Floating Planets (FFPs), or dark and low-mass objects in our Galaxy is only possible through gravitational microlensing \citep{Einstein1936,Liebes1964,1964MNRASrefsdal}. However, the infrared surveys with long exposure times from young stellar clusters led to the discovery several isolated and giant planets \citep[e.g., ][]{2000SciOsorio}. FFPs in our galaxy can bend and magnify the light of background and collinear source stars temporarily during a timescale mostly less than $10$ days. Dense temporal microlensing observations by the Microlensing Observations in Astrophysics (MOA, \citet{moa2001,Sumi2003}) and The Optical Gravitational Lensing Experiment (OGLE, \citet{OGLE_IV,OGLE2003_1}) discovered several FFPs in short-duration microlensing events toward the Galactic bulge \citep{2011NaturSumi,Mroz2017Natur}. Recently, by analyzing data from the MOA-II microlensing survey in $2006$-$2014$, \citet{Sumi2023} found that there are $\sim21$ FFPs or very wide orbit planets with masses in the range $[0.33M_{\oplus},~6660 M_{\oplus}]$ per star. Similarly and based on $2016$-$2019$ KMTNet microlensing data, \citet{2022GouldJKAS} reported an exceed of Galactic Earth-mass and super-Earth-mass FFPs related to microlensing events with the Einstein timescales less than half a day, as was predicted in \citet{Mroz2017Natur}.

Detecting new FFPs in different lines of sight in our galaxy (not only toward the Galactic bulge) is crucial for studying their distributions, abundance, characterizations, and their fraction in the Galactic halo dark mater \citep[see, e.g., ][]{2021sajadian}. Short-duration microlensing events due to FFPs or low-mass lens objects are potentially detectable in the TESS observations by chance. In this work, we study this point. We first calculate the microlensing optical depth, event rate (per star per day), and then estimate the expected number of events (per star) for the TESS CTL stars and the targets in FFIs numerically. To find what kinds of microlensing events can be detected in the TESS data, we perform Monte-Carlo simulations of such events from the TESS stars in the CTL and FFIs. We extract the detectable events and investigate their physical and lensing parameters. In near future and based on these simulations we will search real microlensing events in the public TESS data. Additionally, comparing the results from these simulations and the real ones from investigating the TESS public data helps to test/evaluate the models describing our galaxy (e.g., the mass function for FFPs).

The outline of the paper is as following. In Section \ref{sec1}, we consider the TESS CTL stars as source stars of potential microlensing events and simulate them. We generate synthetic data points taken by TESS. By applying some detectability criteria, we extract the detectable ones and finally explain their results in Subsection \ref{stat}. Then, we consider the targets from the TESS FFIs as the source stars of potential microlensing events and we simulate these events. The details and the results from these simulations are given in Section \ref{FFI}. In the last section, \ref{conclu}, we explain the results and conclusions.

\section{Simulating Microlensing from the TESS CTL}\label{sec1}
Source stars of microlensing events from the TESS observations can be inside either (i) the Candidate Target List (CTL, \citet{2018AJstassun,2019AJstassun}), or (ii)  Full-Frame Images (FFIs, \citet{2018AJFFIstars}). The observing strategies for these two lists are different and hence we study their microlensing events separately as explained in this section and the next one, respectively. To simulate a microlensing event we need to specify a source star, and a lens object. Then, by specifying the lensing parameters we generate the microlensing light curves and their synthetic data points. Finally, we examine if simulated events are detectable or not.

We take $\sim 160,000$ stars from the CTL catalog (as microlensing source stars, \citet{2018AJstassun,2019AJstassun}) based on their priorities from the MAST(Mikulski Archive for Space Telescopes) catalog \citep{ctlTESS}. The information from these source stars which are necessary for making microlensing events have been reported in this catalog. Then, we determine the lens distance from the observer based on the microlensing event rate function which is $\Gamma \propto \rho_{\rm{tot}}(l,~b,~D_{\rm l}) \times \sqrt{x_{\rm{ls}}(1-x_{\rm{ls}})}$, where $x_{\rm{ls}}=D_{\rm l}/D_{\rm s}$, $D_{\rm s}$ is the source distance from the observer, and $D_{\rm l}$ is the lens distance. Here, $\rho_{\rm{tot}}(l,~b,~D_{\rm l})$ is the overall mass density in our Galaxy for a given line of sight versus distance from the observer due to the Galactic bulge, thin disk, thick disk and stellar halo. $(l,~b)$ represent the Galactic longitude and latitude which specify the observing line of sight toward each CTL target. These density distributions are taken from the Besan\c{c}on model\footnote{\url{https://model.obs-besancon.fr/}} \citep{Robin2003,Robin2012}.

The minimum and maximum observing time spans for a given source star by TESS are $27.4$ and $356.2$ days, respectively. We therefore expect both short-duration and long-duration microlensing events to be potentially detectable in its observations. Hence, we choose the lens mass from this wide range $M_{\rm l}\in [0.1 M_{\oplus}, 2M_{\odot}]$, which includes free-floating planets (FFPs), brown dwarfs (BDs), and main-sequence stars (MSs). According to the recent results offered by \citet{Sumi2023}, there are $\sim 21$ FFPs or very wide-orbit planets with masses in the range $[0.33,~6660] M_{\oplus}$ in our Galaxy per star, with the total mass of $\sim 80 M_{\oplus}$ per star. These results were verified in other references and through other microlensing survey observations \citep[e.g., ][]{Mroz2017Natur,2022GouldJKAS,2021McDonald}. Accordingly, in the simulation we use the normalized mass function offered in Figure (6) of \citet{Sumi2023} to determine the mass of the lens objects for this wide range ($M_{\rm l}\in [0.1 M_{\oplus}, 2M_{\odot}]$) as given by:
$$\log_{10}\Big[dN\big/d\log_{10}[M(M_{\oplus})]\Big]=~~~~~~~~~~~~~~~~~~~~~~~~~~\\$$
\begin{numcases}{}
-0.96\log_{10}[M(M_{\oplus})]+1.69&$\log_{10}[M(M_{\oplus})]<2.6$\nonumber\\
+0.58\log_{10}[M(M_{\oplus})]-2.31&$2.6\leq \log_{10}[M(M_{\oplus})]<4.5$\nonumber\\
-0.13\log_{10}[M(M_{\oplus})]+0.87&$4.5\leq \log_{10}[M(M_{\oplus})]<5.5$\nonumber\\ 
-1.32\log_{10}[M(M_{\oplus})]+7.44&otherwise, 
\label{massf}
\end{numcases}

In fact, we choose the mass of lens objects from $dN\big/dM=1\big/M \times dN\big/d\log_{10}[M]$.

\begin{figure}
\centering
\includegraphics[width=0.49\textwidth]{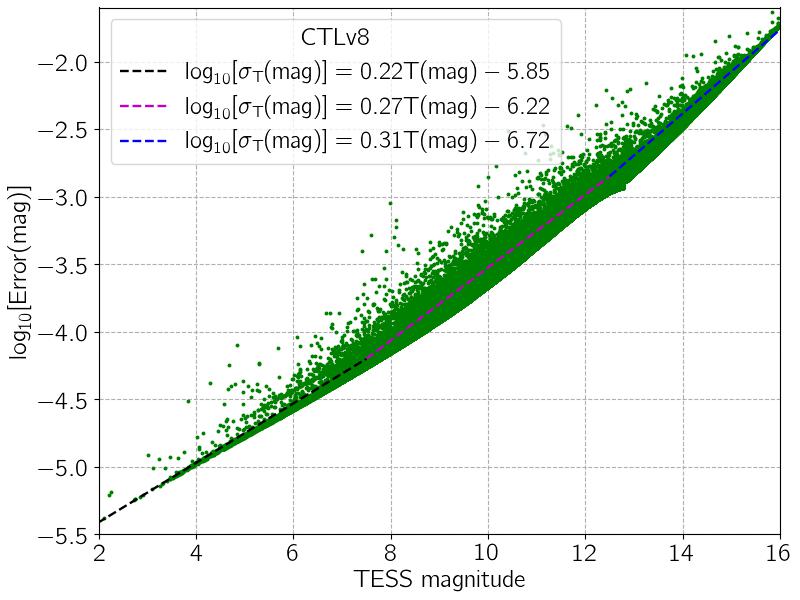}
\caption{The plot shows the median photometric errors versus the apparent magnitudes in the TESS passband $T$ for the TESS CTL targets. Three straight lines are fitted to the data as plotted with the dashed black, magenta and blue lines which reach to each other at the apparent magnitudes $7.5$ and $12.5$ mag.}\label{error}
\end{figure}

\begin{figure*}
\centering
\subfigure[]{\includegraphics[width=0.32\textwidth]{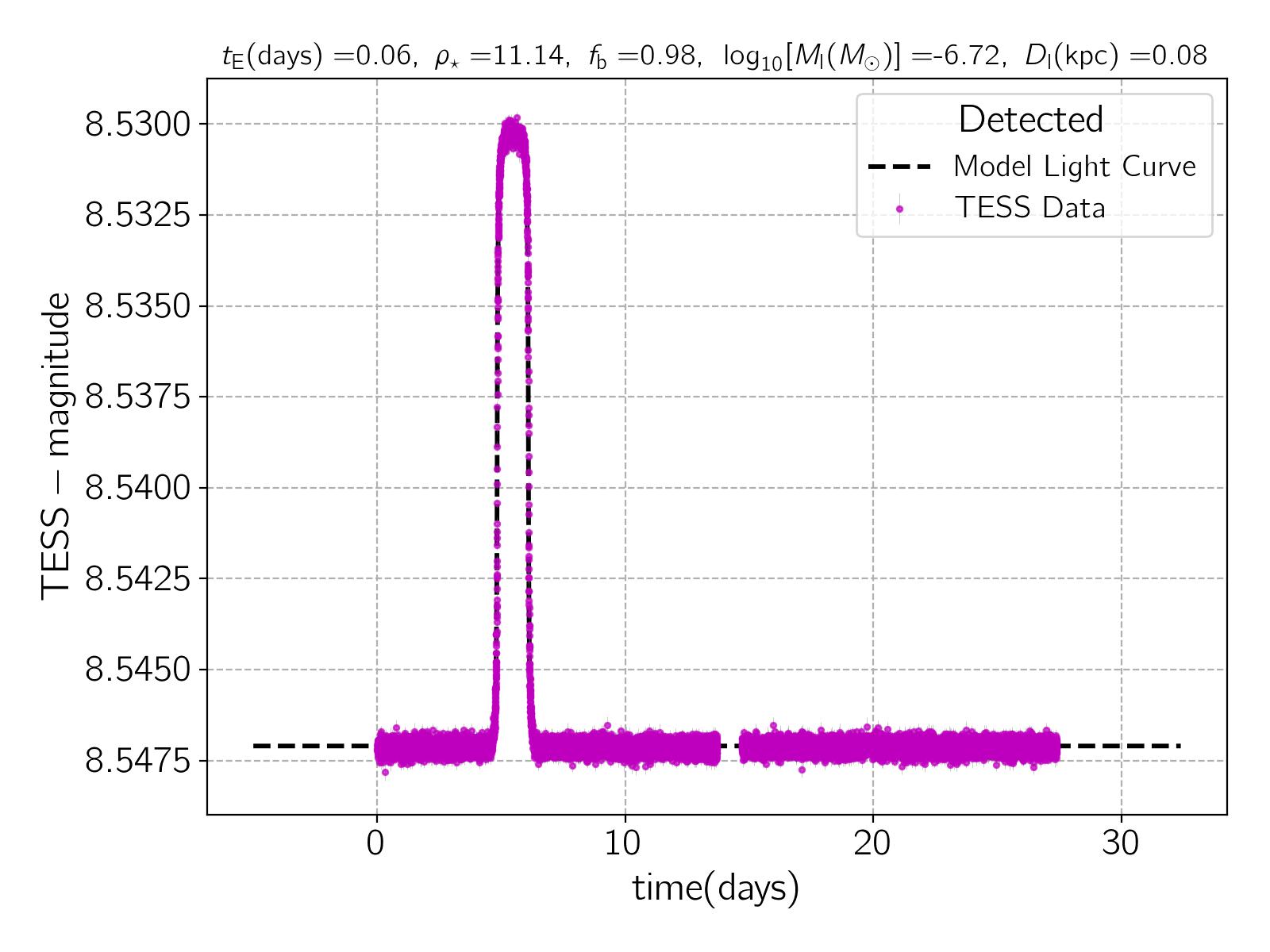}\label{lighta}}
\subfigure[]{\includegraphics[width=0.32\textwidth]{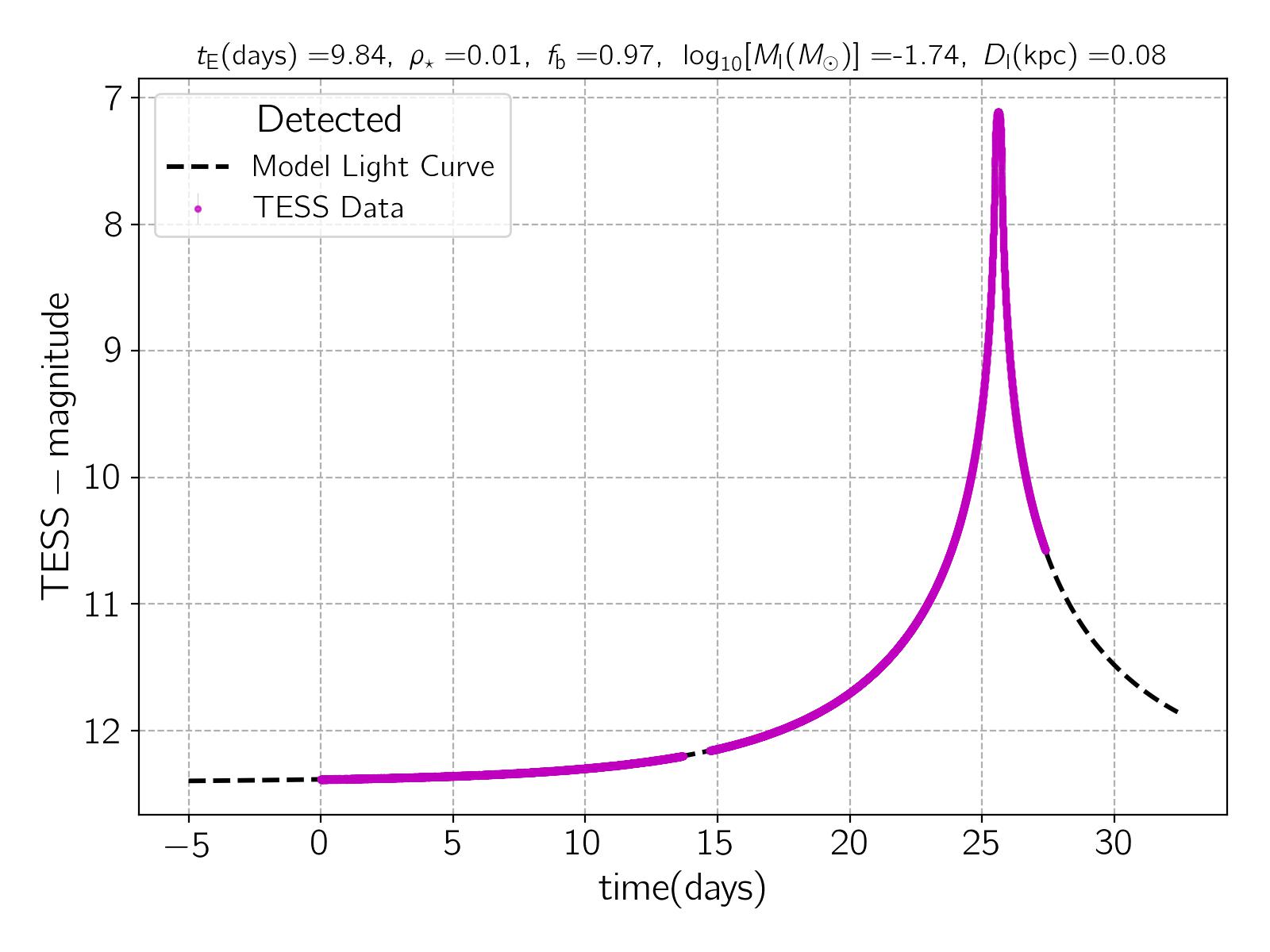}\label{lightb}}
\subfigure[]{\includegraphics[width=0.32\textwidth]{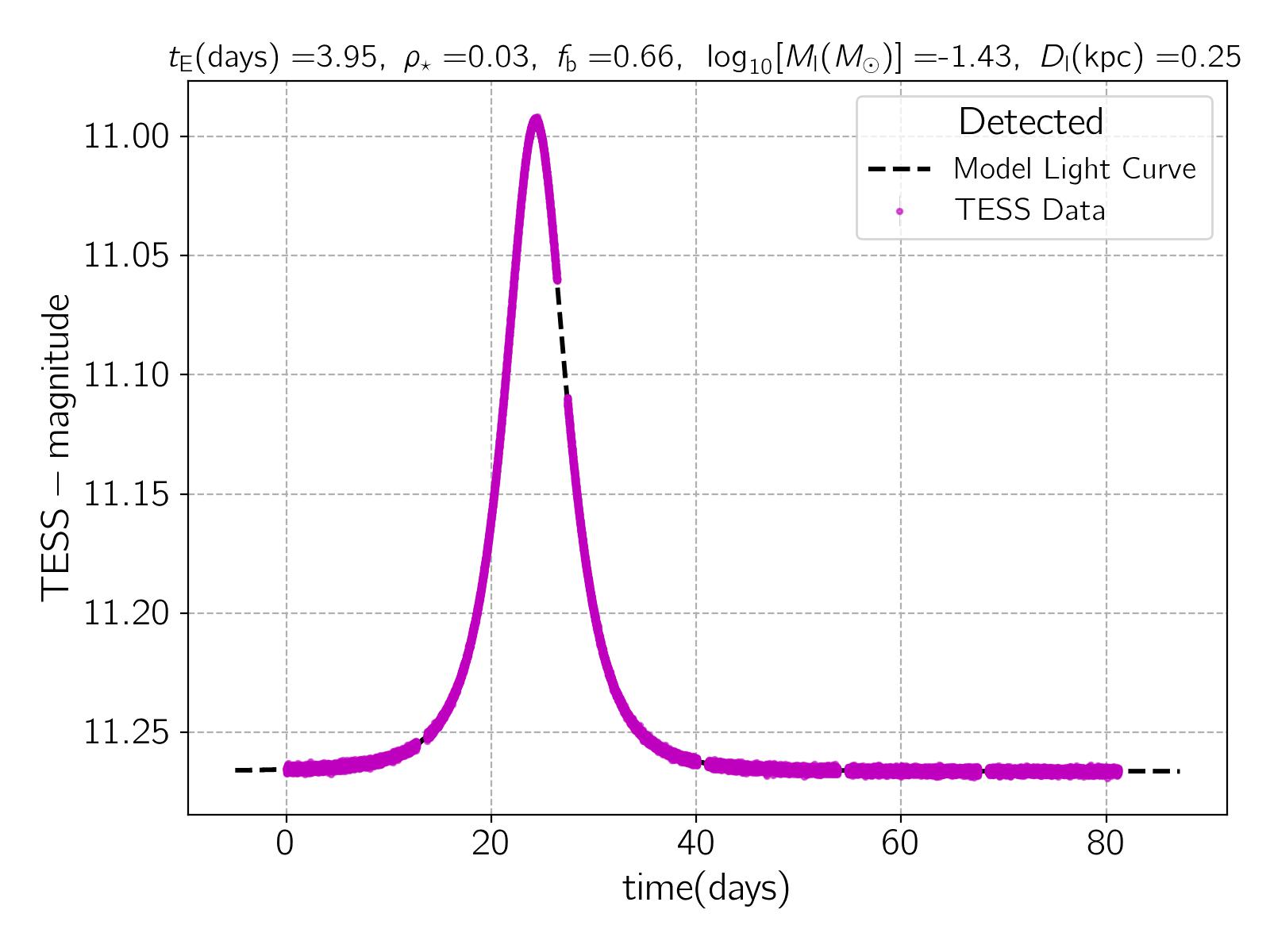}\label{lightc}}
\subfigure[]{\includegraphics[width=0.32\textwidth]{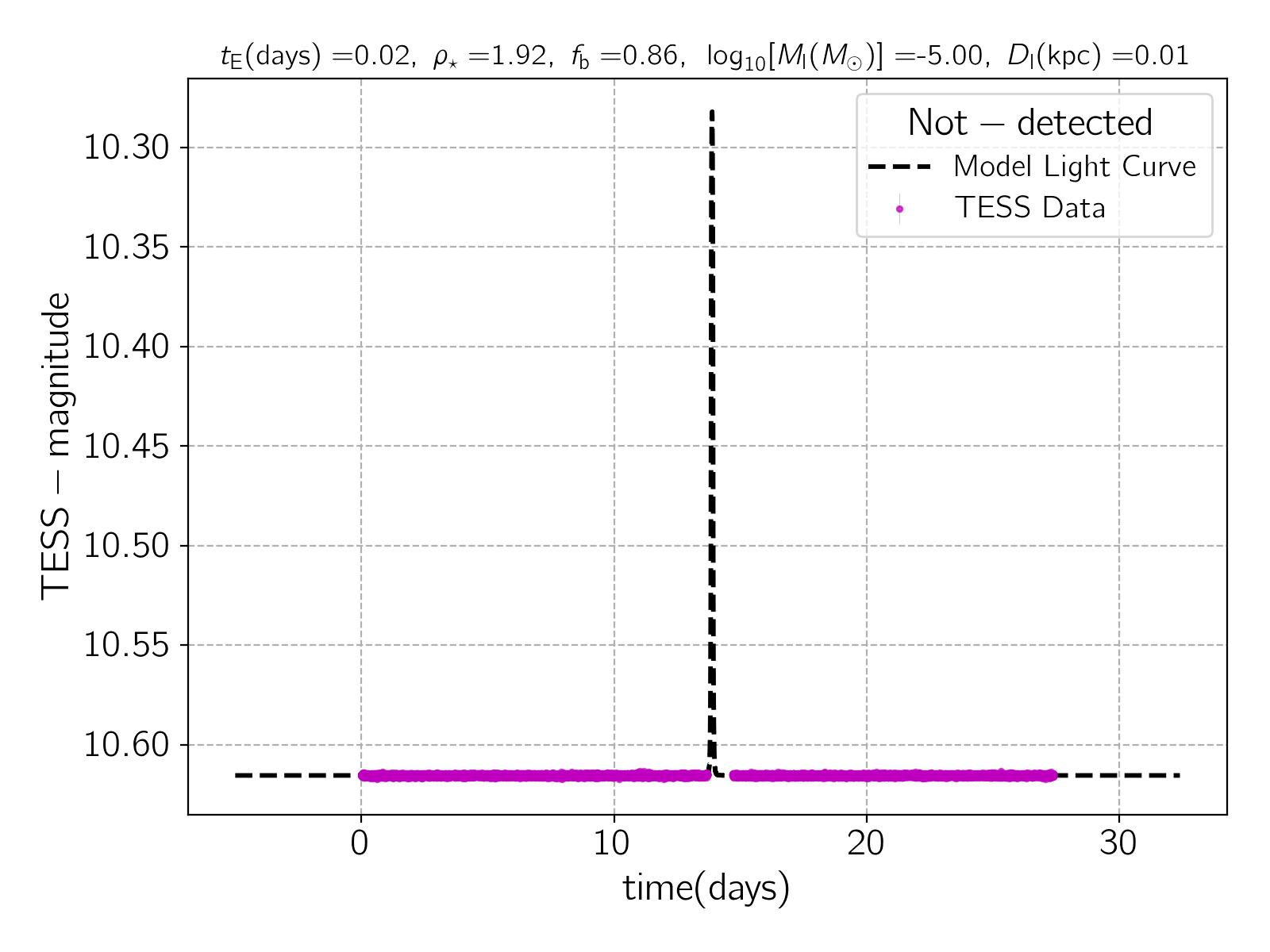}\label{lightd}}
\subfigure[]{\includegraphics[width=0.32\textwidth]{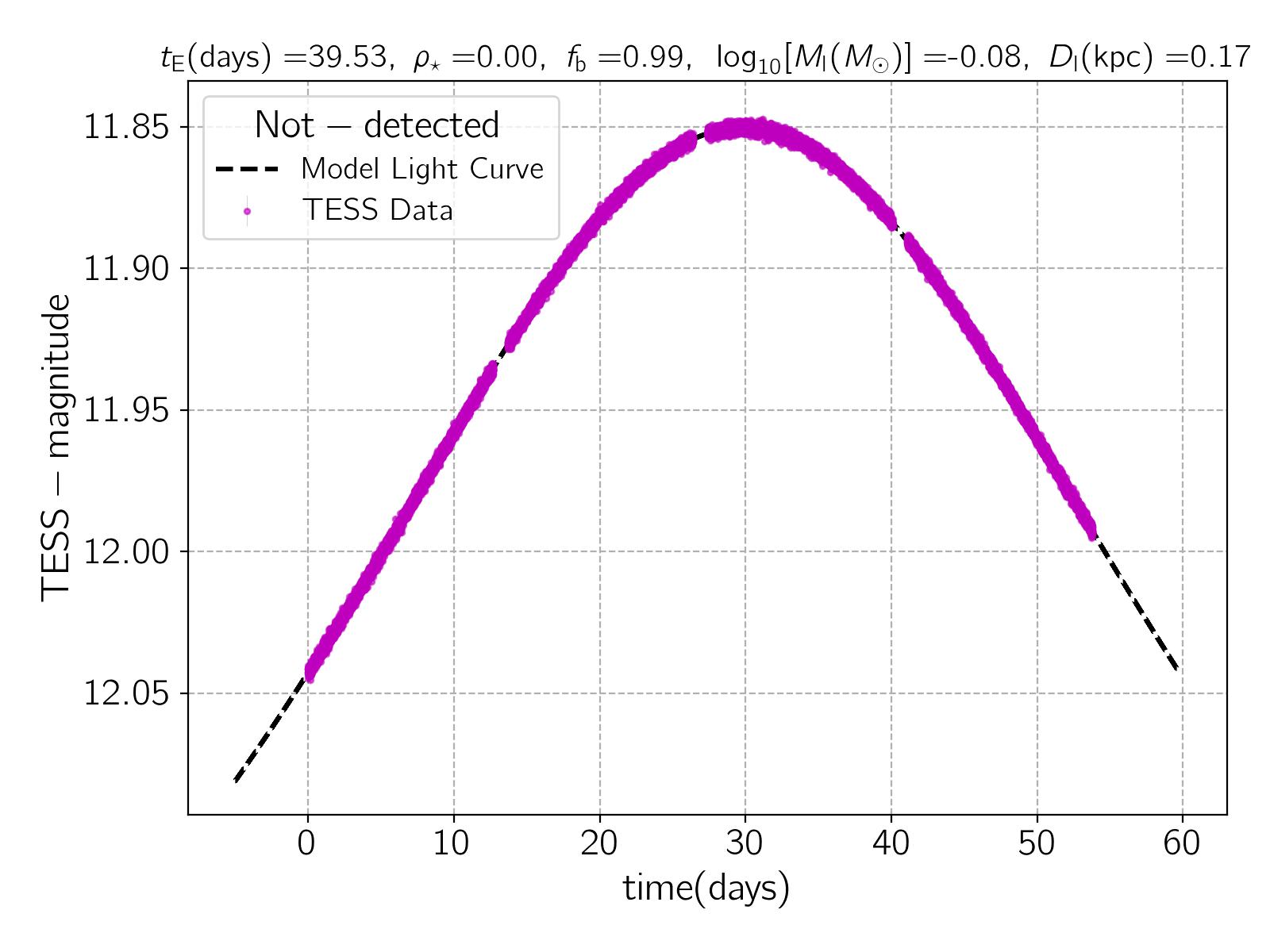}\label{lighte}}
\subfigure[]{\includegraphics[width=0.32\textwidth]{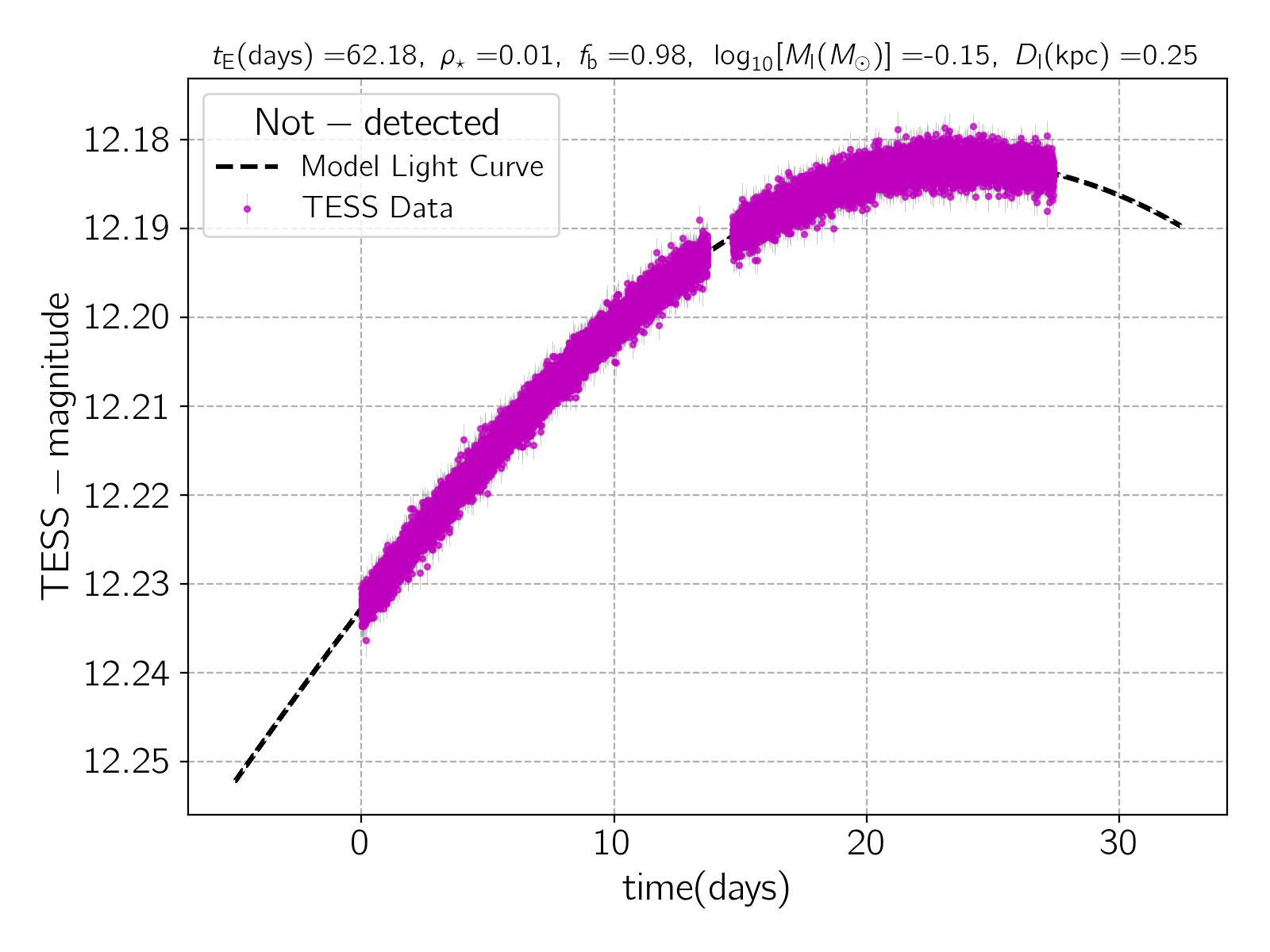}\label{lightf}}
\caption{Three top panels represent examples of detectable microlensing events from the TESS CTL targets, and three bottom panels show simulated microlensing events from the CTL stars which are not detectable according to our detectability criteria. The parameters of each light curve are mentioned at the top of its frame.}\label{light}
\end{figure*}

\begin{figure*}
\centering
\includegraphics[width=0.49\textwidth]{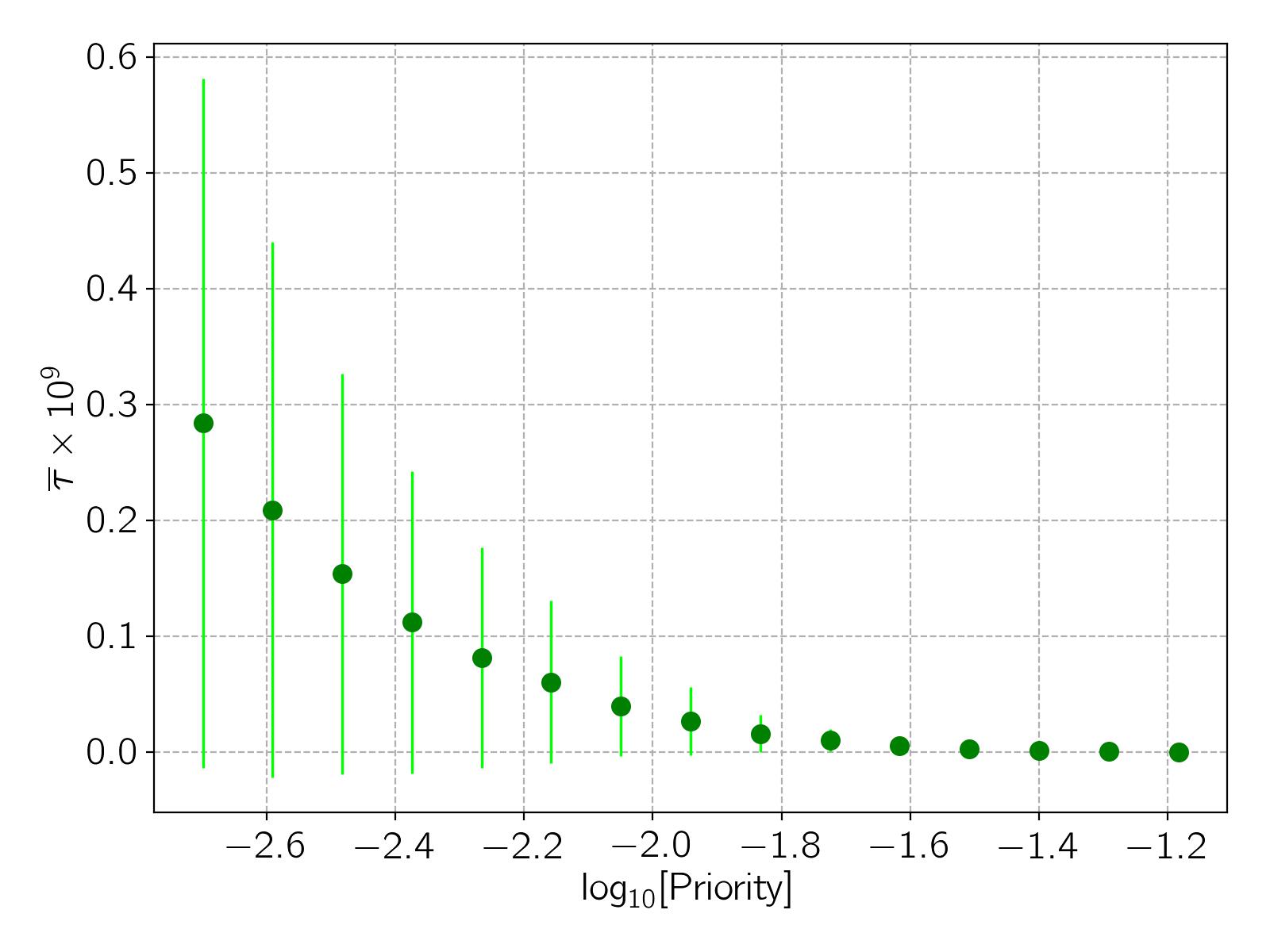}
\includegraphics[width=0.49\textwidth]{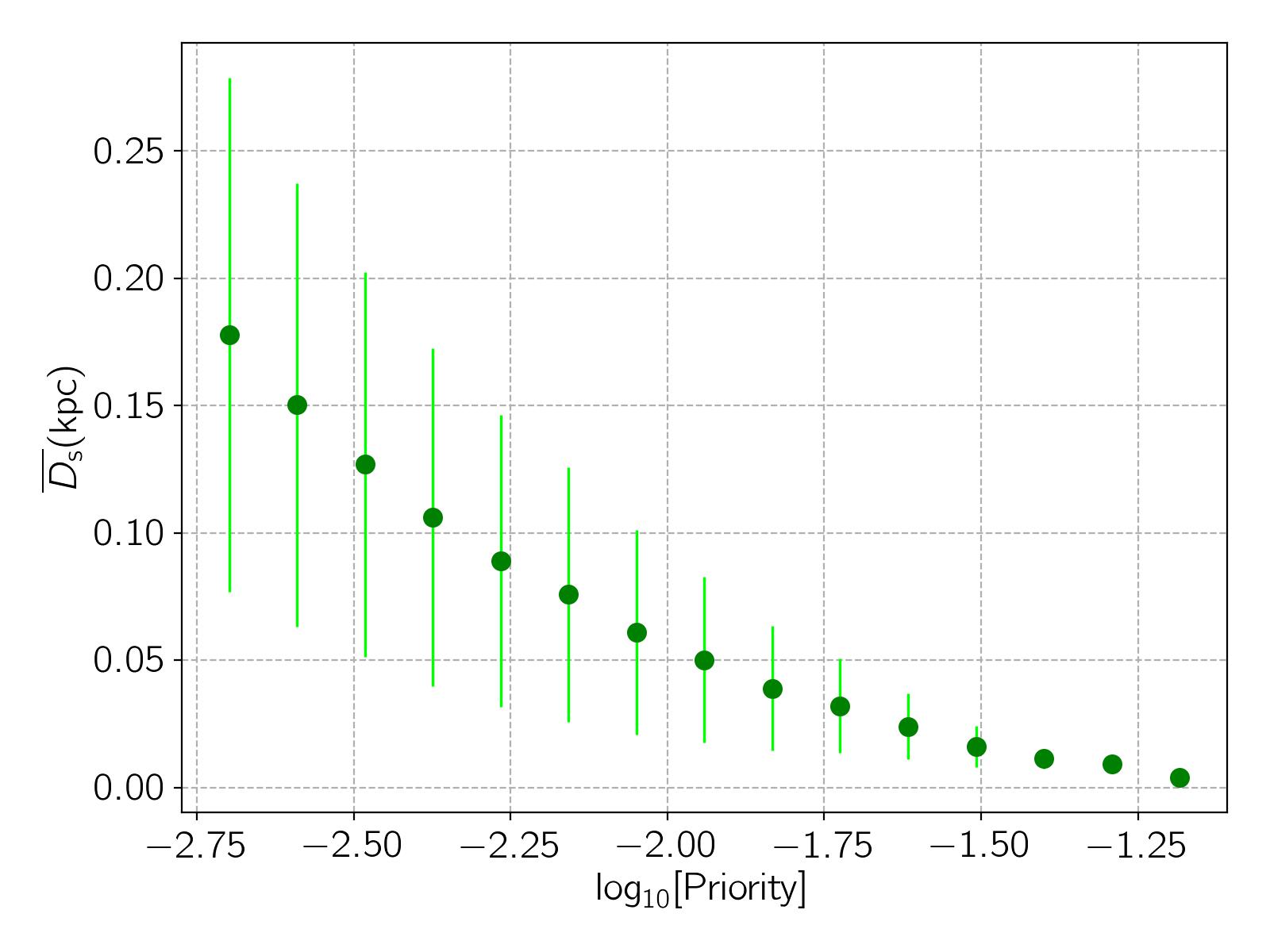}
\caption{Left panel displays the average lensing optical depth (given by Equation \ref{tau}) versus the TESS CTL priority (in the logarithmic scale), and the right panel shows the average source distance $\overline{D_{\rm s}}(\rm{kpc})$ in simulated microlensing events versus the source stars' priority. The error bars represent the standard deviations of these two parameters derived from their samples.}\label{optp}
\end{figure*}

The time of the closest approach of the lens from the source center is uniformly chosen from the range $[0,~T_{\rm{obs}}]$, where $T_{\rm{obs}}$ is the observation window. For the parts of sky covered only by one sector (during two years) this observation window is $27.4$ days. The lens impact parameter is chosen smoothly from the range $[0,~1]$. In most of short-duration microlensing events (specially the ones due to FFPs) the finite-source effect \citep{1994wittmoa} is significant and will change the lensing magnification factors. However, for modeling these short-duration microlensing events with extreme finite-source sizes there is a continuous degeneracy between the Einstein timescale, finite-source size, and the blending factor \citep[see, e.g., ][]{2022ApJ,2023MNRAS.521.6383S}. To incorporate the finite-source effect, we use the RT-model\footnote{\url{http://www.fisica.unisa.it/gravitationastrophysics/VBBinaryLensing.htm}} \citep{2010MNRASBozza,2018MNRASBozza}. 

For each source star taken from the TESS CTL, we calculate the blending factor (the ratio of the source flux to the total baseline flux) using $f_{\rm b}=1/\big(1+ \rm{ContamRatio}\big)$, where 'ContamRatio' is the ratio of contaminating flux to flux from the star \citep[see, Table 9 of ][]{2018AJstassun} which was indicated for all CTL targets.

In the next step, we generate synthetic data points for simulated events according to the considered observing time span $T_{\rm{obs}}$, the cadence, and the photometric error bars. All CTL stars inside a sector do not have the same observational time span which is owing to partially overlapping of sectors. For instance, the number of (the TESS CTL' and FFIs') stars which were observed during a given observing time span were estimated in \citet{2018ApJSBarclay}. Accordingly, $T_{\rm{obs}}$ could be from $27.4$ days (for stars inside only one sector) up to one year of TESS (for stars toward the north and south ecliptic poles).   

The TESS photometric error bars $\sigma_{\rm{m}}$ depend strongly on the stellar apparent magnitudes. We determine these error bars (the squared root of summation of squared stellar noise, count noise, sky noise and readout noise) using the linear best-fitted relations to the observing data for the TESS CTL stars as shown in Figure \ref{error} by different colors. These relations as specified in this figure are
\begin{numcases}{\log_{10}[\sigma_{\rm m}(\rm{mag})]=}
0.22~m_{\rm{T}}-5.85&$m_{\rm{T}} \le 7.5$ \nonumber\\
0.27~m_{\rm{T}}-6.22& $7.5 < m_{\rm{T}}\leq 12.5$ \nonumber\\
0.31~m_{\rm{T}}-6.72&otherwise,
\label{errorT}
\end{numcases}
where $m_{\rm{T}}$ is the stellar apparent magnitude in the TESS passband $T$. However, we add some statistical errors to the values extracted from Eq. \ref{errorT} (in the logarithmic scale) based on a Gaussian function $\mathcal{N}(0,~0.1)$, and finally limit the least photometric errors to $10^{-5}$ mag. The observing cadence for the TESS CTL targets is fixed at $2$ minutes. 

After simulating the synthetic data points, we extract the events which are detectable according to four criteria: (i) at least three data points should be above the baseline by $5 \sigma_{\rm m}$, (ii) $\Delta \chi^{2}>500$, where $\Delta \chi^{2}= |\chi^{2}_{\rm{real}}-\chi^{2}_{\rm{base}}|$ is the difference between $\chi^{2}$ values from fitting the real and baseline models, (iii) there should be at least $5$ data points over the baseline (where the magnification factor is less than $1.1$), and (iv) each side of light curves should be covered by at least $3$ data points. Two last criteria eliminate long-duration events for which the synthetic data points cover only small parts of light curves.   

Some examples of simulated light curves are represented in Figure \ref{light}. The lensing parameters of these microlensing events are mentioned at the top of each frame. The magenta data points are simulated based on the TESS observing strategy for the CTL targets. There are one-day gaps between every $12.7$-day window of observation. The observational time spans for the light curves \ref{lighta}, \ref{lightb}, \ref{lightd}, and \ref{lightf} are $27.4$ days (i.e., their source stars lie inside one sector), and for two others, \ref{lighte}, and \ref{lightc}, are $54.8$, and $82.2$ days, respectively (i.e., their source stars lie inside two and three continuous sectors). In this figure, three bottom events are not detectable through the TESS observations, because either the magnification curve occurs inside the TESS observing gap (\ref{lightd}), or there are no observing data on their baselines \ref{lighte} and \ref{lightf}. The detectable microlensing events (three top panels) are either due to the low-mass lens objects with considerable finite-source size (\ref{lighta}) or high-magnification ones.

For each possible observing time span (which could be $\rm{No}_{\rm s}\times 27.4$ days, where $\rm{No}_{\rm s}$ is the number of sectors overlapping for a given part of sky), we perform a Monte-Carlo simulation of detectable microlensing events from the CTL source stars. We note that two overlapping sectors are after each other where the difference between their ecliptic longitudes is $\simeq 27$ degrees. Hence, for two overlapping sectors we set their observing time $2\times 27.4= 54.8$ days. However, the overlapping part of two sectors 1, 13 (or 14 and 26 in the northern hemisphere) was observed by TESS during two $27.4$-days time spans with a $300$-day gap. In the simulation by ignoring this case, we assume each overlapping part is observed continuously. The results of all simulations are explained in the following subsection.  
  
\begin{figure*}
\centering
\includegraphics[width=0.32\textwidth]{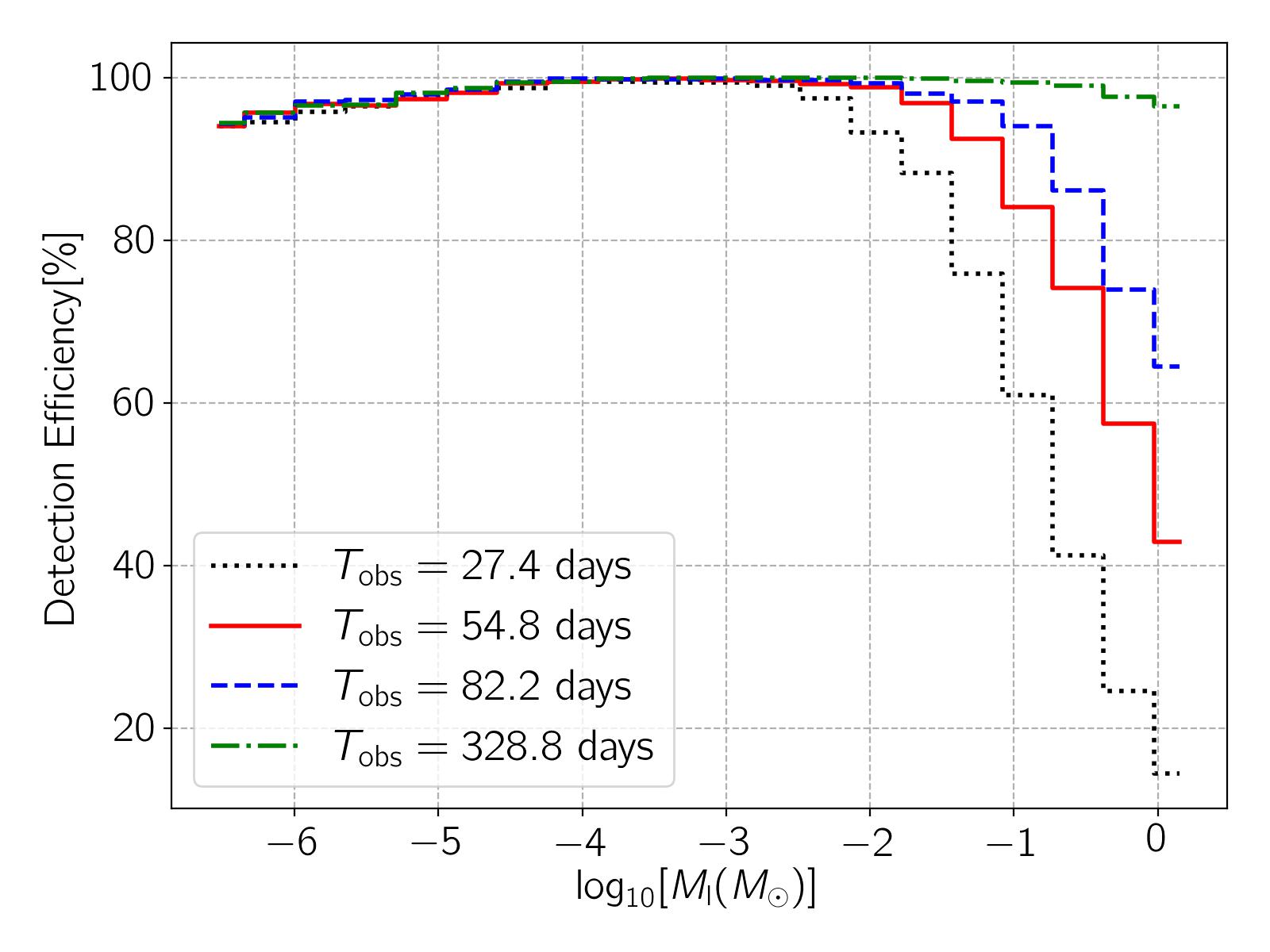}
\includegraphics[width=0.32\textwidth]{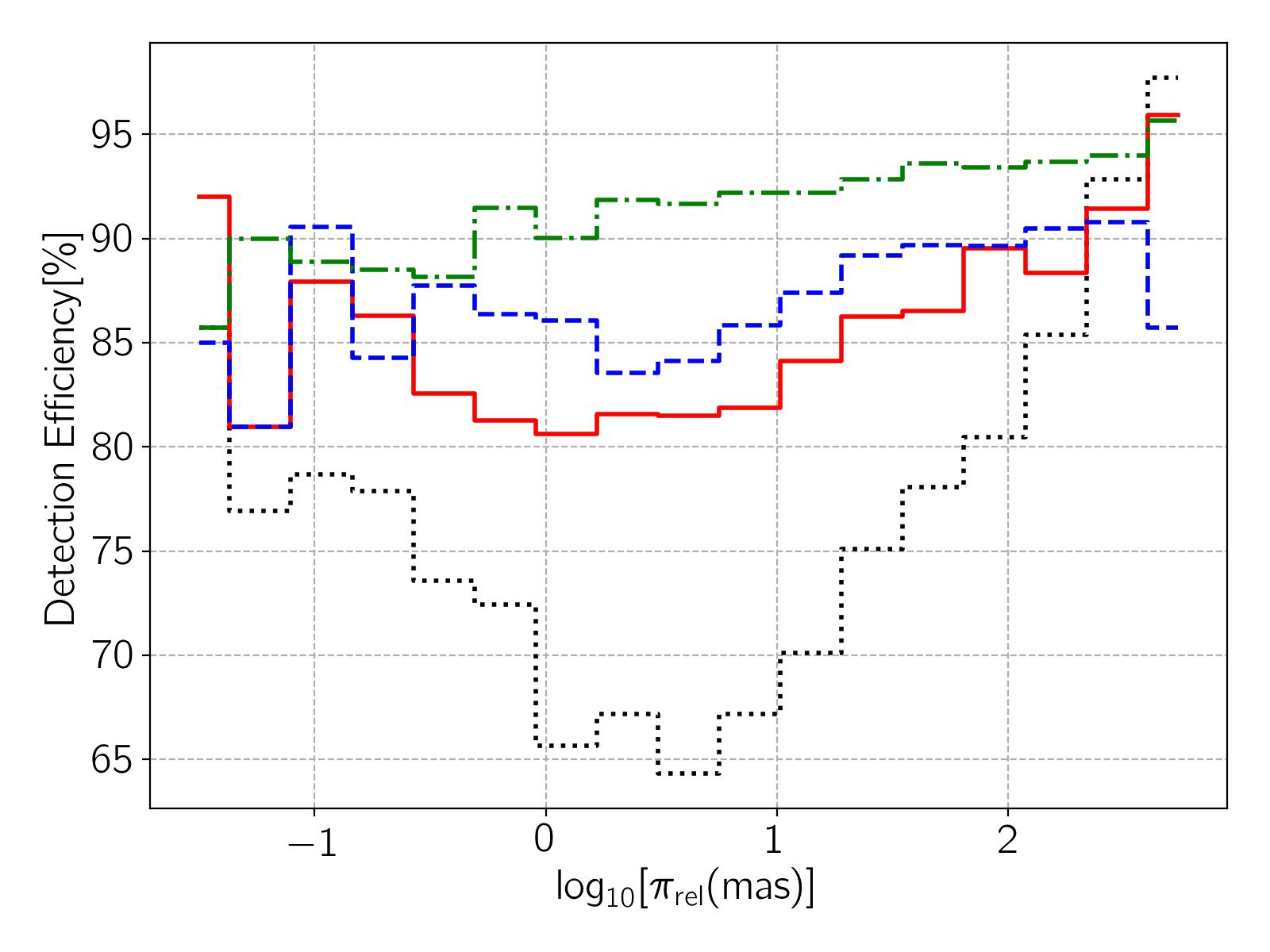}
\includegraphics[width=0.32\textwidth]{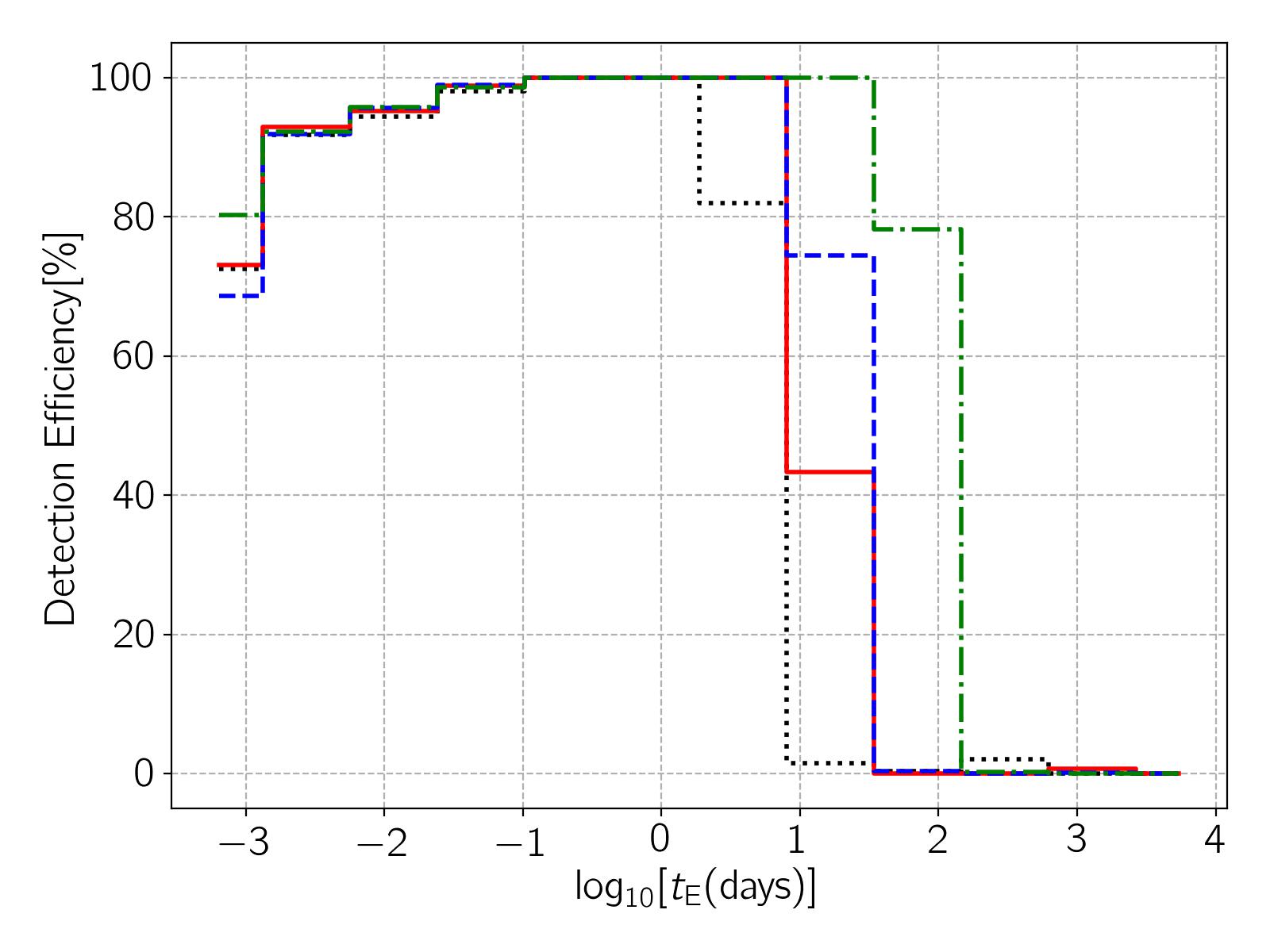}
\includegraphics[width=0.32\textwidth]{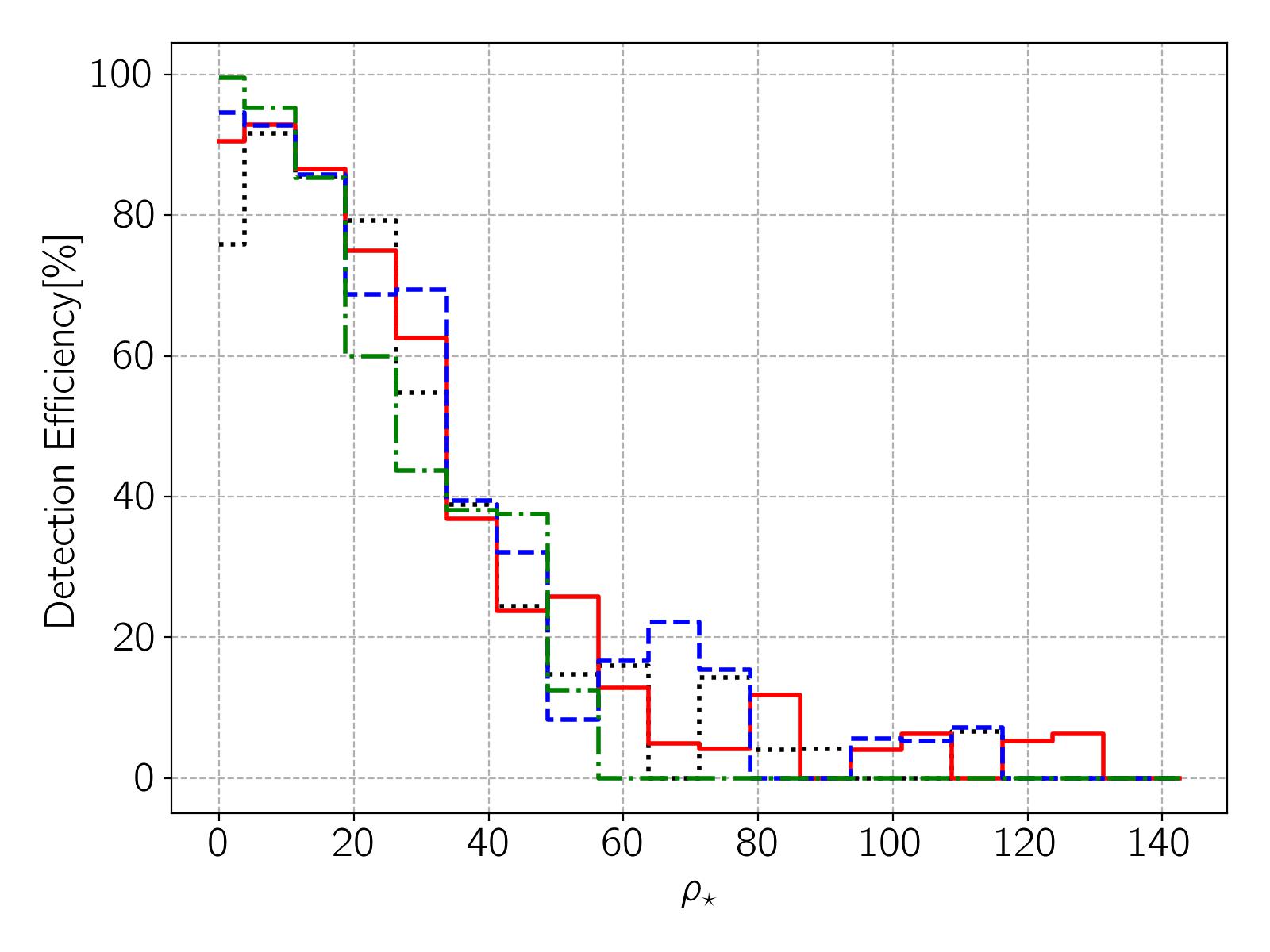}
\includegraphics[width=0.32\textwidth]{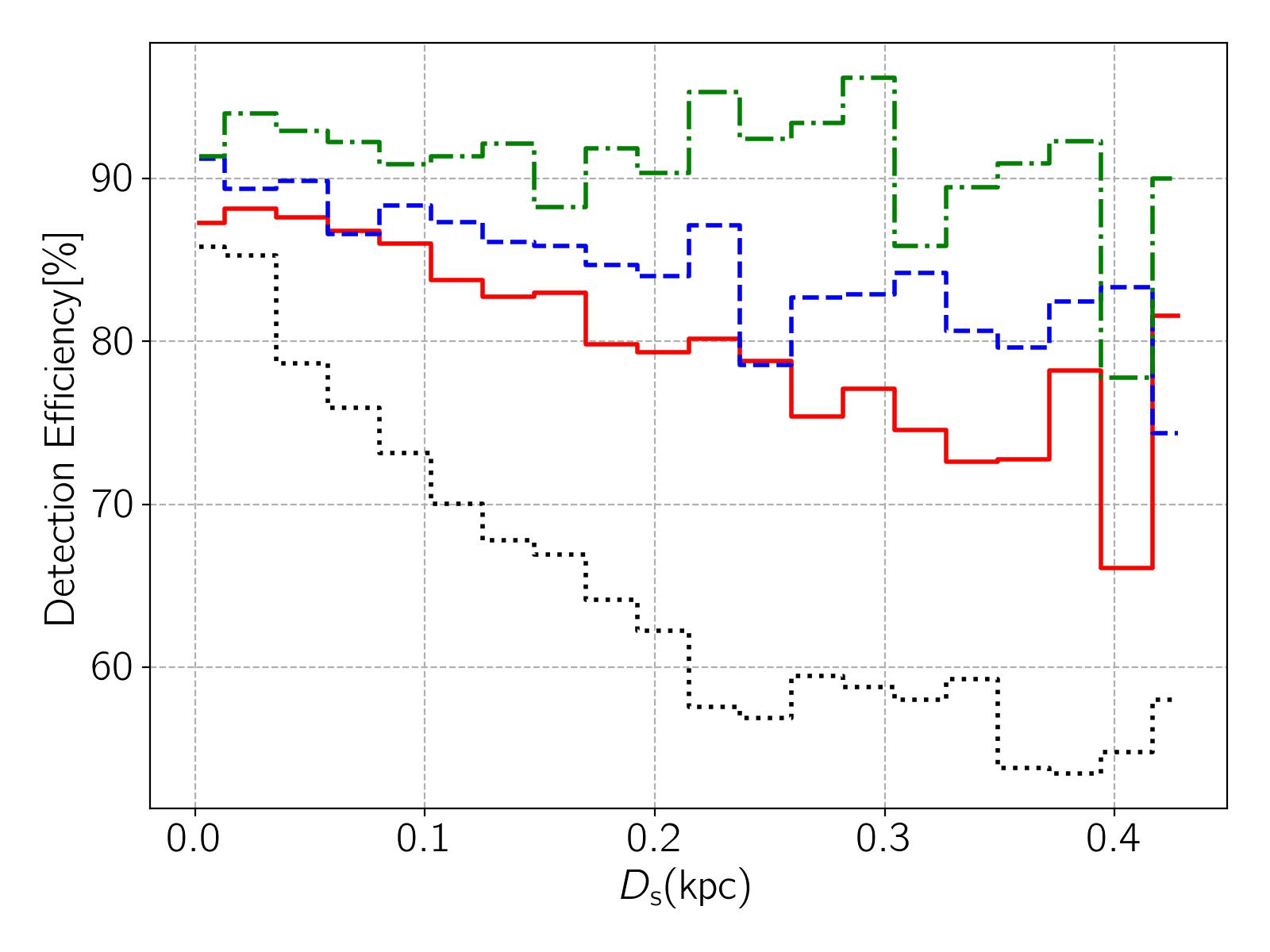}
\includegraphics[width=0.32\textwidth]{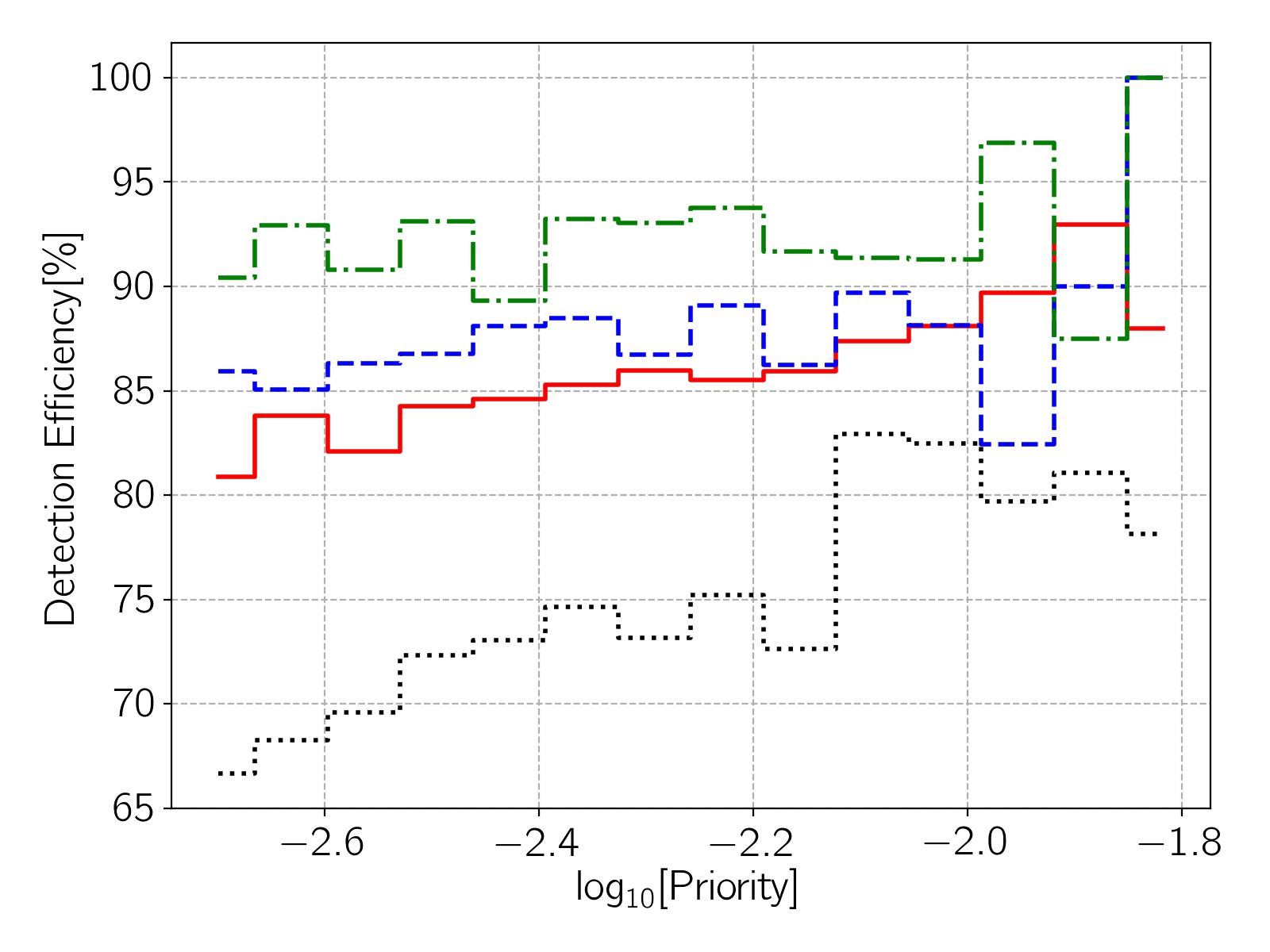}
\caption{These panels show the efficiencies (i.e., the ratio of simulated and detectable microlensing events to total simulated events) for detecting microlensing events versus six parameters, including the lens mass $M_{\rm l}(M_{\odot})$,  the relative parallax amplitude in the logarithmic scale $\log_{10}[\pi_{\rm{rel}}]$, the Einstein crossing time in the logarithmic scale $\log_{10}[t_{\rm E}(\rm{days})]$, the projected source radius normalized to the Einstein radius $\rho_{\star}$, the source distance $D_{\rm s}(\rm{kpc})$, and the CTL priority in the logarithmic scale, respectively, in percent. The different line styles and colors represent different observing time spans which are mentioned in the first panel.}\label{effi}
\end{figure*}

\subsection{Results from Monte Carlo simulations}\label{stat}
The TESS CTL targets have been (and are) observed based on their priority. So, we first study the correlation between the lensing optical depth and the priority. The lensing optical depth shows the probability of occurring a lensing event which is given by \citep[see, e.g.,  ][]{1986ApJPaczynski}:  
\begin{eqnarray}
\tau=\frac{4\pi G~D_{\rm s}^{2}}{c^{2}} \int_{0}^{1}  \sum_{i=1}^{4}\rho_{i}(l, b, D_{\rm{\rm l}}) x_{\rm{ls}}(1-x_{\rm{ls}}) dx_{\rm{ls}}.
\label{tau}
\end{eqnarray}
This formula calculates the ratio of the cumulated angular areas covered by the angular Einstein rings around any possible lensing object which are collinear with a given source star to the considered angular area around that source star. Here, $G$ is the gravitational constant, and $c$ is the light velocity. The summation is done over the mass densities, $\rho_{i}$s, due to different structures in our galaxy including the Galactic bulge, thin and thick disks, and stellar halo. The left panel of Figure \ref{optp} shows the average optical depth (Eq. \ref{tau}) versus the CTL priority (in the logarithmic scale) as determined in the MAST catalog. Hence, by decreasing the priority the lensing optical depth increases. The TESS CTL targets with higher priorities are on average closer to the observer, as shown in the right panel of Figure \ref{optp}. For closer stars to the observer the lensing probability is less than that for farther ones. 

\begin{deluxetable*}{c c c c c c c c c c}
\tablecolumns{10}
\centering
\tablewidth{0.9\textwidth}\tabletypesize\footnotesize
\tablecaption{The lower and upper quartile values (i.e., $Q_{1}$, $Q_{3}$ which are $25$th and $75$th percentiles) of the lensing parameters' distributions of simulated and detectable microlensing events from the TESS CTL stars. \label{tab1}}
\tablehead{\colhead{$\rm{No}_{\rm s}$}&\colhead{$D_{\rm l}$}&\colhead{$D_{\rm s}$}&\colhead{$\log_{10}[M_{\rm l}]$}&\colhead{$\log_{10}[t_{\rm E}]$}&\colhead{$\log_{10}[R_{\rm E}]$}&\colhead{$v_{\rm{rel}}$}&\colhead{$\rho_{\star}$}&\colhead{$\log_{10}[\pi_{\rm E}]$}&\colhead{$m_{\rm{TESS}}$}\\
& $\rm{kpc}$&$\rm{kpc}$&$M_{\odot}$& $\rm{days}$ & $\rm{au}$ & $\rm{km}/s$ & & &$\rm{mag}$}
\startdata
$1$ & $0.03,~0.10$ & $0.08,~0.19$ & $-6.39,~-5.88$ & $-1.93,~-1.38$ & $-3.61,~-3.26$ & $18.8,~47.1$ & $2.4,~8.3$ & $2.81,~3.32$ & $9.3,~11.6$\\
$2$ & $0.03,~0.10$ & $0.08,~0.19$ & $-6.39,~-5.88$ & $-1.94,~-1.37$ & $-3.61,~-3.26$ & $18.9,~47.5$ & $2.4,~8.2$ & $2.81,~3.32$ & $9.2,~11.5$\\
$3$ & $0.03,~0.10$ & $0.08,~0.19$ & $-6.39,~-5.87$ & $-1.92,~-1.36$ & $-3.61,~-3.26$ & $18.5,~46.8$ & $2.4,~8.5$ & $2.80,~3.32$ & $9.3,~11.5$\\
$4$ & $0.03,~0.10$ & $0.08,~0.19$ & $-6.37,~-5.85$ & $-1.91,~-1.36$ & $-3.60,~-3.25$ & $18.7,~45.9$ & $2.4,~8.3$ & $2.80,~3.31$ & $9.2,~11.6$\\
$5$ & $0.03,~0.10$ & $0.08,~0.19$ & $-6.39,~-5.87$ & $-1.93,~-1.37$ & $-3.60,~-3.26$ & $18.8,~47.6$ & $2.4,~8.3$ & $2.80,~3.31$ & $9.3,~11.5$\\
$6$ & $0.03,~0.10$ & $0.08,~0.19$ & $-6.38,~-5.87$ & $-1.95,~-1.36$ & $-3.61,~-3.26$ & $18.9,~48.2$ & $2.4,~8.2$ & $2.80,~3.31$ & $9.2,~11.5$\\
$7$ & $0.03,~0.10$ & $0.08,~0.19$ & $-6.39,~-5.87$ & $-1.94,~-1.35$ & $-3.61,~-3.26$ & $18.5,~47.7$ & $2.4,~8.6$ & $2.80,~3.30$ & $9.3,~11.5$\\
$8$ & $0.03,~0.10$ & $0.08,~0.19$ & $-6.39,~-5.86$ & $-1.93,~-1.37$ & $-3.60,~-3.26$ & $18.6,~47.2$ & $2.3,~8.1$ & $2.80,~3.31$ & $9.3,~11.6$\\
$9$ & $0.03,~0.10$ & $0.08,~0.19$ & $-6.39,~-5.86$ & $-1.92,~-1.36$ & $-3.60,~-3.26$ & $18.8,~47.0$ & $2.3,~8.3$ & $2.79,~3.31$ & $9.3,~11.5$\\
$10$ & $0.03,~0.10$ & $0.08,~0.19$ & $-6.38,~-5.86$ & $-1.92,~-1.37$ & $-3.61,~-3.26$ & $18.5,~47.4$ & $2.4,~8.4$ & $2.80,~3.31$ & $9.3,~11.5$\\
$11$ & $0.03,~0.10$ & $0.08,~0.19$ & $-6.39,~-5.86$ & $-1.93,~-1.35$ & $-3.60,~-3.26$ & $18.7,~47.3$ & $2.4,~8.4$ & $2.80,~3.30$ & $9.2,~11.6$\\
$12$ & $0.03,~0.09$ & $0.08,~0.19$ & $-6.39,~-5.88$ & $-1.94,~-1.37$ & $-3.60,~-3.27$ & $18.8,~47.5$ & $2.3,~8.2$ & $2.81,~3.33$ & $9.3,~11.6$\\
$13$ & $0.03,~0.10$ & $0.08,~0.19$ & $-6.38,~-5.85$ & $-1.92,~-1.35$ & $-3.59,~-3.25$ & $18.3,~47.3$ & $2.4,~8.2$ & $2.80,~3.30$ & $9.3,~11.6$\\
\enddata
\tablecomments{Each row reperecents the TESS observing time which is $\rm{No}_{\rm s} \times 27.4$, where No$_{\rm s}$ is the number of sectors covering a given part of sky.}
\end{deluxetable*}
In Figure \ref{effi}, we show the detection efficiency (i.e., the ratio of detectable microlensing events to the total simulated ones) in percent versus six parameters. These parameters are the lens mass in the logarithmic scale $\log_{10}[M_{\rm l}(M_{\odot})]$, the relative parallax amplitude in the logarithmic scale $\log_{10}[\pi_{\rm{rel}}(\rm{mas})]$, the Einstein timescale in the logarithmic scale $\log_{10}[t_{\rm E}(\rm{days})]$, the source radius projected on the lens plane and normalized to the Einstein radius (representing the finite-source size) $\rho_{\star}$, the source distance $D_{\rm s}(\rm{kpc})$, and the CTL priority in the logarithmic scale, respectively. In these plots the different line styles and colors show the detection efficiencies from different Monte-Carlo simulations (by considering different observing times). We note that the parallax effect does not affect detectability of simulated events. Nevertheless, it can change the lensing and physical parameters of microlenses extracted from modeling with respect to their real values for short-duration microlensing light curves\citep{2024sangtarash}. We remind that the parallax amplitude is the relative parallax amplitude normalized to the angular Einstein radius, as following 
\begin{eqnarray}
\pi_{\rm E}=\frac{\pi_{\rm{rel}}}{\theta_{\rm E}}, ~~~~\theta_{\rm E}=\sqrt{\kappa~M_{\rm l}(M_{\odot})~\pi_{\rm{rel}}},
\label{para}
\end{eqnarray} 
where $\pi_{\rm{rel}} (\rm{mas})=1\big/D_{\rm l}(\rm{kpc})-1\big/ D_{\rm s} (\rm{kpc})$ the so-called relative parallax, and $\kappa=8.13~(\rm{mas}\big/M_{\odot})$ is a constant. We note that the Einstein radius is $R_{\rm E}=\theta_{\rm E}\times D_{\rm l}$. According to these plots, we list some noticeable points in the following.  
\begin{itemize}[leftmargin=2.0mm]
\item The highest detection efficiency occurs for the lens objects with the masses $\log_{10}[M_{\rm l} (M_{\odot})] \in [-4.5, -2.5]$, which means the TESS telescope captures the lensing signals for the CTL targets due to super-Earth up to Jupiter-mass FFPs with the efficiency $\sim 100\%$. This point shows the importance of searching for microlensing events in the TESS public data which potentially leads to detect nearby FFPs in all possible lines of sight in our galaxy.

\item When the lens objects have very low masses, e.g., $M_{\rm l}\lesssim10^{-6}M_{\odot}$, they make very short-duration events with considerable parallax and finite-source effects, which are barely detectable in the TESS data. Additionally, massive lens objects can make detectable events only if the lens objects are either very close to or far from the observer (which results either very large or very small relative parallax amplitudes, respectively). For that reason, the detection efficiency for massive lens objects (e.g., main-sequence stars) is not zero, although it decreases with the lens mass. Hence, if massive lens objects are either very close to or very far from the observer, they make detectable events which result considerable detection efficiencies for  $\pi_{\rm{rel}}\ll 0.1$ mas and $\pi_{\rm{rel}}\gg100$ mas (see the second panel of Figure \ref{effi}).

\item The sensitivity of the TESS observations to the Einstein timescales depend on the observing duration, as depicted in the third panel of Figure \ref{effi}. We note that the detection efficiency is $\sim 100\%$ when $\log_{10}[t_{\rm E}(\rm{days})]\in [-0.9,~0.1]$ by considering any observing strategy.

\item The events with larger finite-source sizes have less magnification factors and accordingly less detection efficiency. 

\item Although the lensing optical depth decreases by reducing the source distance, the detection efficiency is higher for closer source stars. The closer source stars are brighter with on average lower photometric errors (see Figure \ref{error}). 

\item The last panel shows that the events from source stars with higher priorities have higher detection efficiency although these source stars are on average closer (see the right panel of Figure \ref{optp}). In fact, the CTL stars with higher priorities are brighter with lower photometric errors.     
\end{itemize}
\begin{figure*}[t]
\centering
\includegraphics[width=0.49\textwidth]{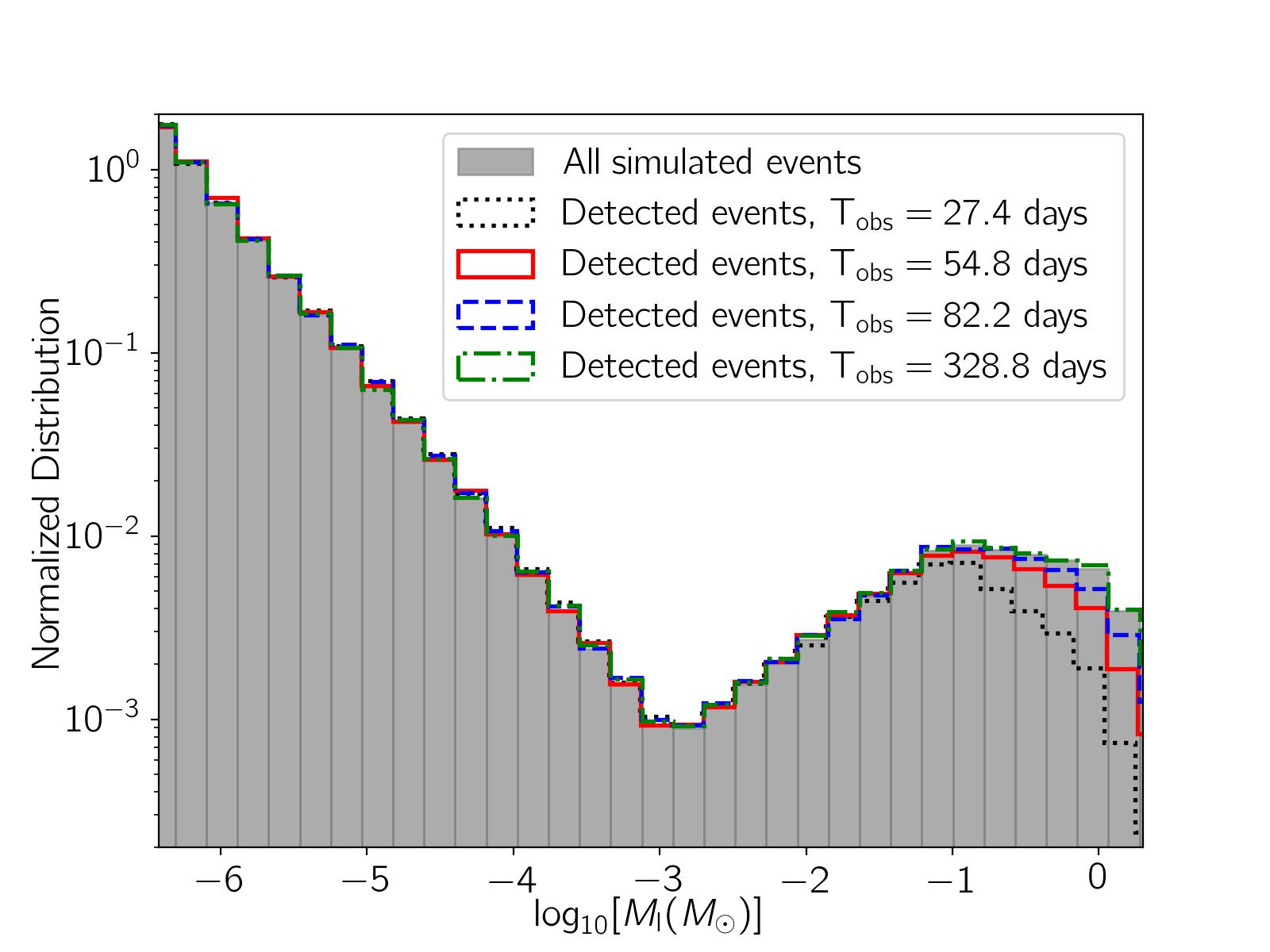}
\includegraphics[width=0.49\textwidth]{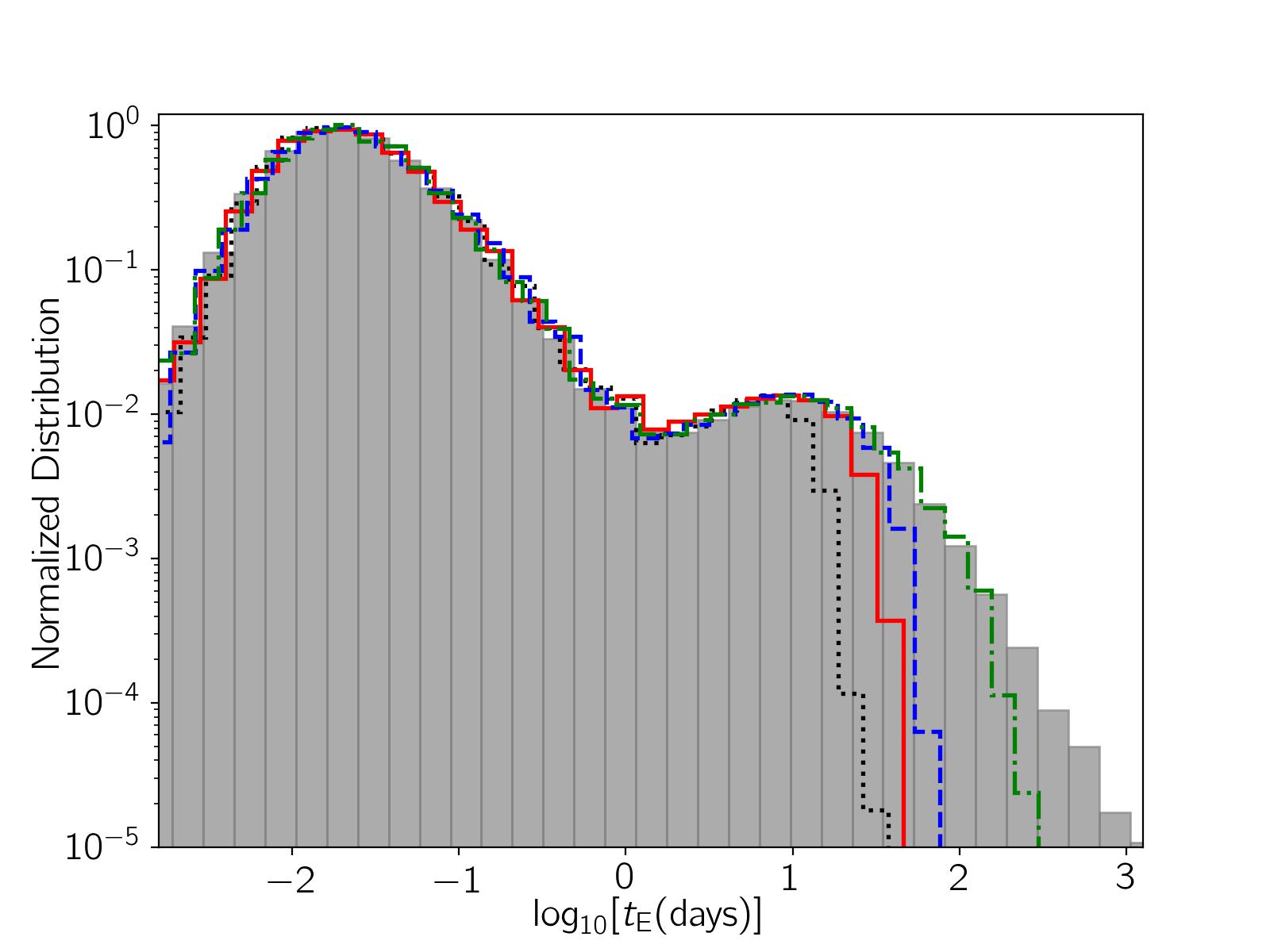}
\includegraphics[width=0.49\textwidth]{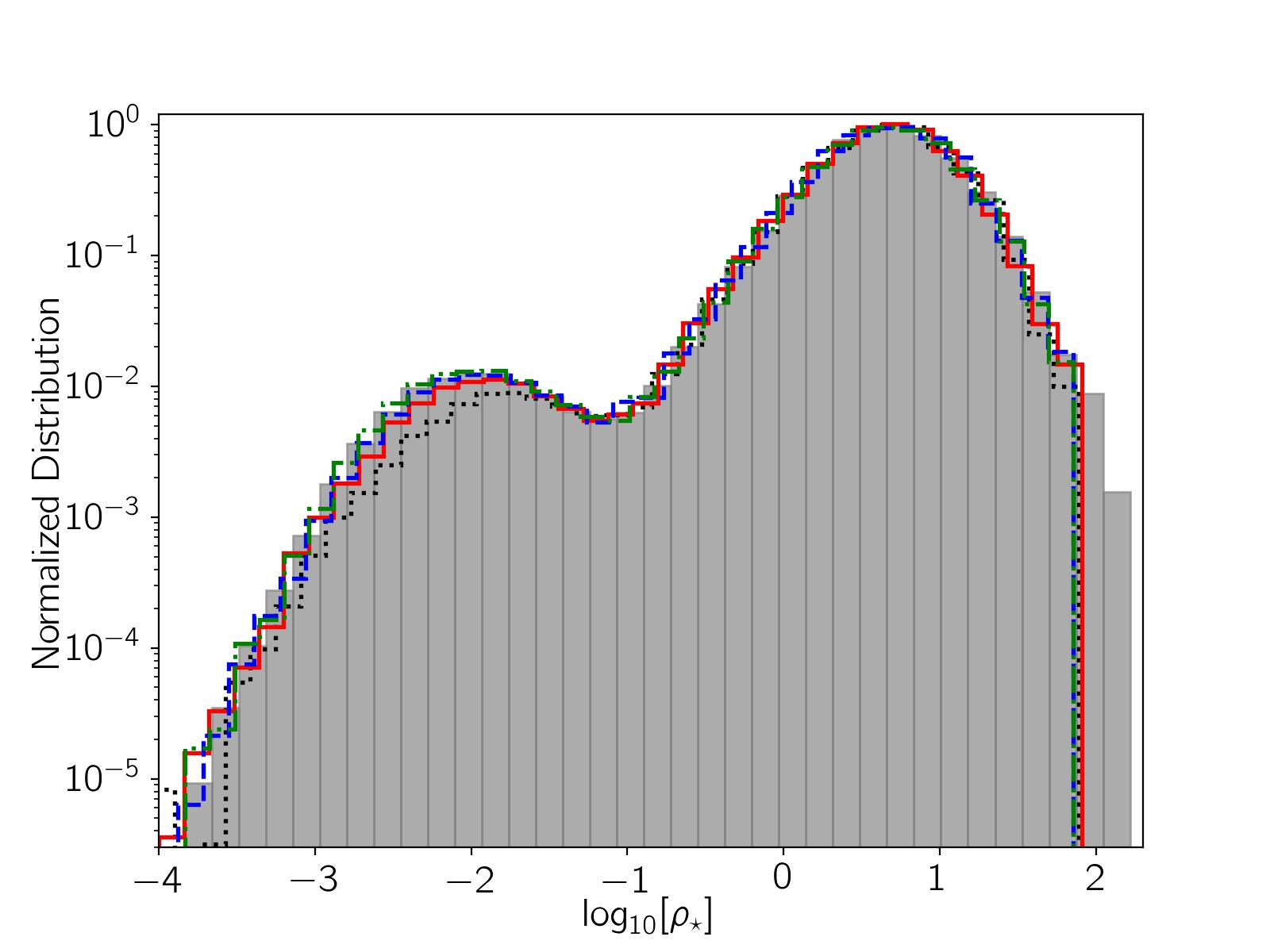}
\includegraphics[width=0.49\textwidth]{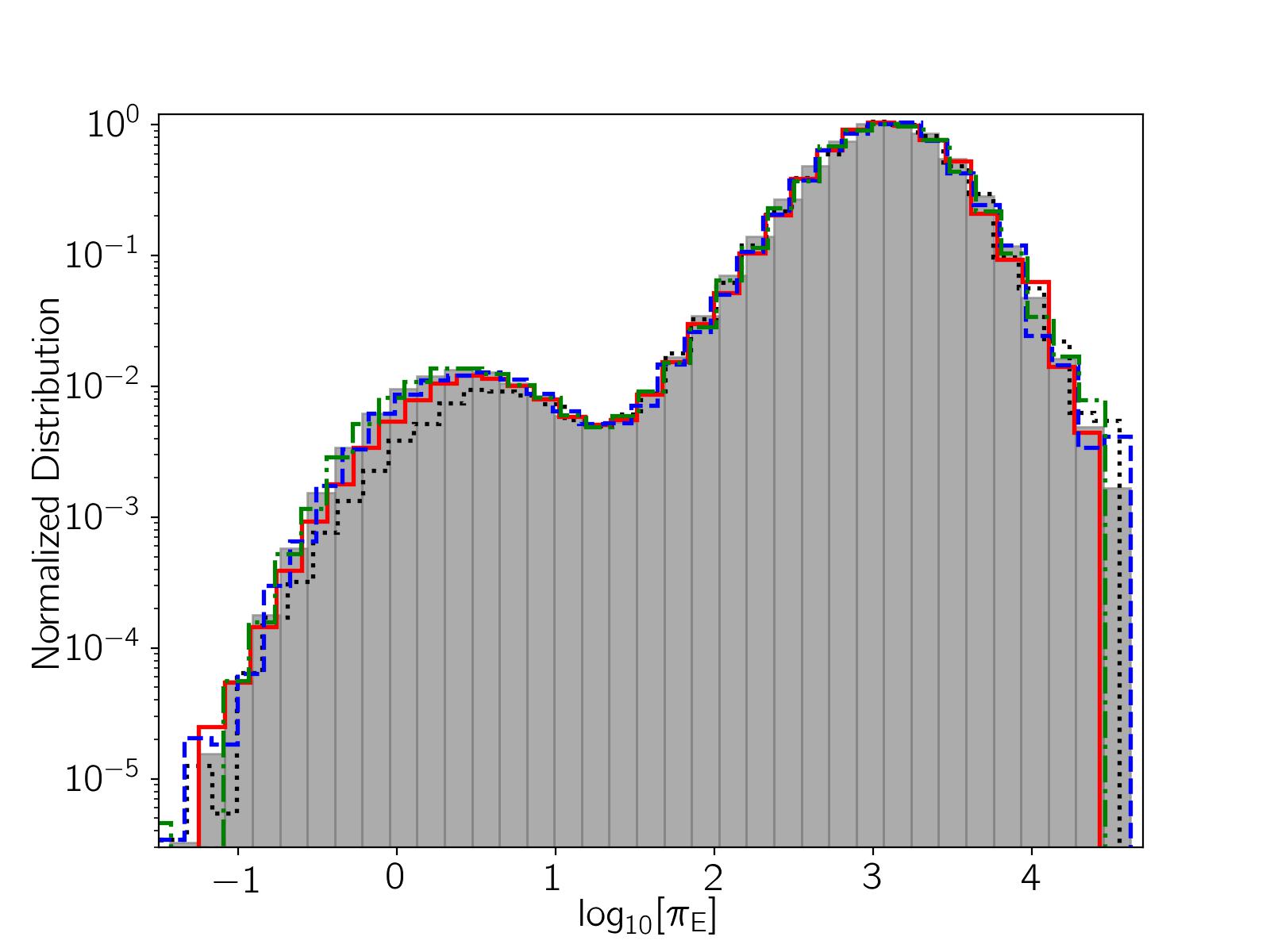}
\caption{The panels show the normalized distributions of four lensing parameters due to all simulated events (gray filled), detectable events by applying different observing times as specified in the first panel. We note that the area under each histogram is normalized to one.}\label{histo}
\end{figure*}

In Table \ref{tab1}, we report the lower and upper quartile values of the lensing parameters due to detectable microlensing events which are extracted from simulations. In this table, every row refers to different observing durations which are corresponding to numbers of sectors covering a given part of the sky ($T_{\rm{obs}}=\rm{No}_{\rm s} \times 27.4$ days, where $\rm{No}_{\rm s}$ is the number of overlapping sectors). Because of the wide range for the lens mass (from low-mass FFPs to MS stars) applied in the simulation resulting in wide possible ranges of $t_{\rm E}$, $\rho_{\star}$, and $\pi_{\rm E}$, the average values of these parameters are not meaningful. We therefore report lower and upper quartiles of these parameters. We show the unit-normalized underlying distributions from simulated events for four mentioned parameters (in  the logarithmic scale) in Figure \ref{histo} with gray color which emphasize their wide ranges. In this figure, their unit-normalized underlying distributions from detectable events by considering different observing time spans are represented with step lines. The first panel shows $\log_{10}\big[dN/d\log_{10}[M_{\rm l}(M_{\odot})]\big]$ as given in Equation \ref{massf}. However, in the simulation we choose the lens mass from $dN/dM_{\rm l}$.

The statistical parameters extracted from simulations are mentioned in Table \ref{tab2}. We first calculate the microlensing event rate in the limit of $100\%$ efficiency, reported in the fourth column of Table \ref{tab2}, as following:
\begin{eqnarray}
\Gamma (\rm{star}^{-1}. \rm{day}^{-1})= \frac{2}{\pi}~\frac{\overline{\tau}}{\left<t_{\rm E}(\rm{days})\right>},
\end{eqnarray}
where, the average for the lensing optical depth $\overline{\tau}$ is done over the optical depths of all simulated events. $\left<t_{\rm E}\right>$ is the relative velocity-weighted average of the Einstein timescale over all simulated events, as given by \citep[e.g.,  ][]{1986ApJPaczynski,2020Spechtmnras}:
\begin{eqnarray}
\left<t_{\rm E}(\rm{days})\right>=\frac{\sum_{i} v_{\rm{rel},~i}~t_{\rm E,~i}(\rm{days})}{\sum_{j}  v_{\rm{rel},~j}}
\end{eqnarray}
where, the summations are done over all simulated microlensing events, so that the timescale and lens-source relative velocity for $i$th event are $t_{\rm E,~i}$, and $v_{\rm{rel},~i}$, respectively. We then evaluate the microlensing event rate for the TESS observation as: 
\begin{eqnarray}
\Gamma_{\rm{TESS}}(\rm{\rm{star}^{-1}. \rm{day}^{-1}})=\Gamma \mathcal{A}=\Gamma\sum_{k=1}^{N_{\rm{bin}}} \epsilon (t_{\rm E, k})~\mathcal{F}(t_{\rm E, k}),
\end{eqnarray} 
where, $\mathcal{A}$ is the factor which determines the average of the TESS detection efficiency over the Einstein timescales, $\epsilon(t_{\rm E, k})$ is the TESS detection efficiency for the Einstein timescale $t_{\rm E, k}$ (as shown in the third panel of Figure \ref{effi}), and $\mathcal{F}(t_{\rm E, k})$ is the unit-normalized underlying distribution of the Einstein timescale (as shown in the second panel of Figure \ref{histo}). Here, the summation is done over different bins for Einstein timescales. We report $\left<t_{\rm E}(\rm{days})\right>$, $\mathcal{A}$, and $\Gamma_{\rm{TESS}}(\rm{\rm{star}^{-1}. \rm{day}^{-1}})$ in the fifth, sixth, and seventh column of Table \ref{tab2}, respectively.


The number of detectable microlensing events per star, reported in the eighth column of Table \ref{tab2}, is derived by using $\hat{N_{\rm e}}(\rm{star}^{-1})=\Gamma_{\rm{TESS}}\times T_{\rm{obs}}$. To have a sense about the number of detectable microlensing events, we take the total number of the CTL stars in sectors 1 to 77 that have been observed or it is planned to be observed in future which is $N_{\rm{tot}}=1,390,486$ \citep{ctlTESS}. To estimate the numbers of these stars which are observed with a given observing duration we use the data in Fig. (2) of \citet{2018ApJSBarclay}. According to this paper, the fractions of stars lie in $\rm{No}_{\rm s}=1$, $2$, $3$, $4$, ..., $11$, $12$, $13$ sectors simultaneously are $f=74.2\%$, $18.6\%$, $2.8\%$, $0.7\%$,  $0.4\%$, $0.4\%$  $0.2\%$,  $0.2\%$, $0.1\%$, $0.1\%$, $0.1\%$, $1.2\%$, and $1.0\%$, respectively. By estimating the number of the CTL stars we evaluate the number of events $N_{\rm{e}}=\hat{N_{\rm e}}(\rm{star}^{-1}) \times N_{\star}$, which are reported in the tenth column of Table \ref{tab2}. Here $N_{\star}=N_{\rm{tot}}\times f$. The $\epsilon$ (the eleventh column) is the TESS efficiency for detecting potential microlensing events (the ratio of detectable events to the total simulated ones) in percent. The last column specifies the fractions of different types (MSs, BDs, FFPs respectively) of lens objects in detectable events. Some noticeable points from these tables are listed in the following. 

\begin{deluxetable*}{c c c c c c c c c c c c}
\tablecolumns{12}
\centering
\tablewidth{0.99\textwidth}\tabletypesize\footnotesize
\tablecaption{Tha average statistical parameters due to simulated microlensing events from the TESS CTL stars. \label{tab2}}    
\tablehead{\colhead{$\rm{No}_{\rm s}$}&\colhead{$T_{\rm{obs}}$}&\colhead{$\overline{\tau} \times 10^{9}$}&\colhead{$\Gamma \times 10^{9}$}&\colhead{$\left<t_{\rm E}\right>$}&\colhead{$\mathcal{A}$}& \colhead{$\Gamma_{\rm{TESS}}\times 10^{9}$}&\colhead{$\hat{N_{\rm e}}\times 10^{8}$}&\colhead{$N_{\star}^{\dag}$}&\colhead{$N_{\rm e}$}&\colhead{$\epsilon$}& \colhead{$f_{\rm{MS}}:f_{\rm{BD}}:f_{\rm{FFP}}$}\\ 
 &$\rm{days}$& &$\rm{star}^{-1}\rm{days}^{-1}$&$\rm{days}$& & $\rm{star}^{-1}\rm{days}^{-1}$& $\rm{star}^{-1}$ & & & $[\%]$ & $[\%]$}
\startdata
$1$ & $27.4$ & $0.21$ & $0.61$ & $0.21$ & $0.94$ & $0.58$ & $1.58$ & $1031 k$ & $0.016$ & $93.8$ & $0.5:0.3:99.2$\\
$2$ & $54.8$ & $0.20$ & $0.59$ & $0.21$ & $0.95$ & $0.56$ & $3.07$ & $258 k$ & $0.008$ & $94.8$ & $0.8:0.3:98.8$\\
$3$ & $82.2$ & $0.20$ & $0.59$ & $0.22$ & $0.95$ & $0.56$ & $4.61$ & $38 k$ & $0.002$ & $95.1$ & $1.0:0.4:98.7$\\
$4$ & $109.6$ & $0.20$ & $0.56$ & $0.23$ & $0.95$ & $0.54$ & $5.87$ & $9 k$ & $0.001$ & $94.9$ & $1.1:0.4:98.6$\\
$5$ & $137.0$ & $0.21$ & $0.62$ & $0.21$ & $0.96$ & $0.59$ & $8.06$ & $5 k$ & $0.000$ & $95.5$ & $1.0:0.4:98.6$\\
$6$ & $164.4$ & $0.20$ & $0.59$ & $0.21$ & $0.96$ & $0.57$ & $9.33$ & $5 k$ & $0.000$ & $95.9$ & $1.1:0.4:98.6$\\
$7$ & $191.8$ & $0.20$ & $0.59$ & $0.22$ & $0.95$ & $0.56$ & $10.80$ & $3 k$ & $0.000$ & $95.1$ & $1.1:0.4:98.5$\\
$8$ & $219.2$ & $0.20$ & $0.60$ & $0.22$ & $0.95$ & $0.57$ & $12.47$ & $2 k$ & $0.000$ & $95.3$ & $1.1:0.4:98.5$\\
$9$ & $246.6$ & $0.21$ & $0.60$ & $0.22$ & $0.95$ & $0.57$ & $13.99$ & $1 k$ & $0.000$ & $95.3$ & $1.1:0.4:98.5$\\
$10$ & $274.0$ & $0.20$ & $0.58$ & $0.22$ & $0.95$ & $0.55$ & $15.09$ & $1 k$ & $0.000$ & $95.1$ & $1.1:0.4:98.5$\\
$11$ & $301.4$ & $0.20$ & $0.59$ & $0.22$ & $0.95$ & $0.56$ & $16.85$ & $1 k$ & $0.000$ & $95.3$ & $1.1:0.4:98.5$\\
$12$ & $328.8$ & $0.20$ & $0.59$ & $0.22$ & $0.95$ & $0.56$ & $18.45$ & $16 k$ & $0.003$ & $95.3$ & $1.1:0.4:98.5$\\
$13$ & $356.2$ & $0.21$ & $0.60$ & $0.22$ & $0.96$ & $0.57$ & $20.28$ & $13 k$ & $0.003$ & $95.5$ & $1.2:0.4:98.5$\\
\enddata
\tablecomments{$^{\dag}$The total number of TESS CTL stars have been observed or planned to be observed is $1,390,486$. We determine their fractions which were (or are) detected by the TESS telescope for each observing time span according to Fig (2) of \citet{2018ApJSBarclay}.}
\end{deluxetable*}

\begin{itemize}[leftmargin=2.0mm]
\item The detectable microlensing events from the TESS CTLs have considerable parallax effect with the parallax amplitude mostly in the range $\log_{10}[\pi_{\rm{E}}] \simeq 2.8$-$3.3$ which is higher than the parallax amplitudes due to common events toward the Galactic bulge ($\pi_{\rm E}\sim 0.3$). According to Equation \ref{para}, $\pi_{\rm E}\propto 1/\sqrt{D_{\rm s} M_{\rm l}}$. Comparing two events one toward the Galactic bulge ($D_{\rm s}\sim 8$ kpc, and $M_{\rm l}\sim 0.3 M_{\odot}$) and another one toward the Galactic disk due to FFPs with $D_{\rm s}\sim 0.2$ kpc and $M_{\rm l} \sim 10^{-7}M_{\odot}$, the ratio of their parallax amplitudes is $\simeq 10000$. Although the parallax effects in such microlensing events are considerable, the parallax-induced deviations in short-duration microlensing light curves being covered by one observing data set (taken by the TESS telescope) are not recognizable. 
	
\item The microlensing events due to the TESS CTLs have considerable $\rho_{\star}$s in the range $\simeq 2$-$9$. Because, $\rho_{\star}\propto 1/\sqrt{D_{\rm s} M_{\rm l}}$. 
	
\item Toward the Galactic disk and for close source stars, the lens-source relative velocity is $\simeq 19$-$48 \rm{km}/s$, which is much lower than that toward the Galactic bulge ($\simeq 100$-$150~\rm{km}/s$).

\item The lensing optical depth for microlensing events of the TESS CTL stars is $0.2\times \sim 10^{-9}$, i.e., three orders of magnitude less than the optical depth toward the Galactic bulge. Its reason is the close distance of the TESS CTL stars to the observer. The resulting event rate is $\sim 10^{-10}$-$10^{-9}$ per star per observing day. Hence, microlensing events for CTL stars happen very rare, but if they happen, they will be very likely to be detected in the TESS light curves.   

\item The total expected number of microlensing events is $N_{\rm{e}, \rm{tot}}\sim 0.03$, i.e., the chance for detecting microlensing events from the TESS observations of the CTL targets (based on our detectability criteria) is too low. 
	
\item We note that $\sim 99\%$ of detected events are due to FFPs in the Galactic disk which are very close to the observer. 
\end{itemize}

In addition to the CTL targets, the FFIs and their stellar light curves taken by the TESS telescope with the cadences $30$, $10$, and $3.3$ minutes were released which are other source star candidates for potential microlensing events. In the next section, we explain the details and results of similar Monte Carlo simulations from the FFI stars as microlensing source stars. 
\begin{figure*}
\centering
\includegraphics[width=0.49\textwidth]{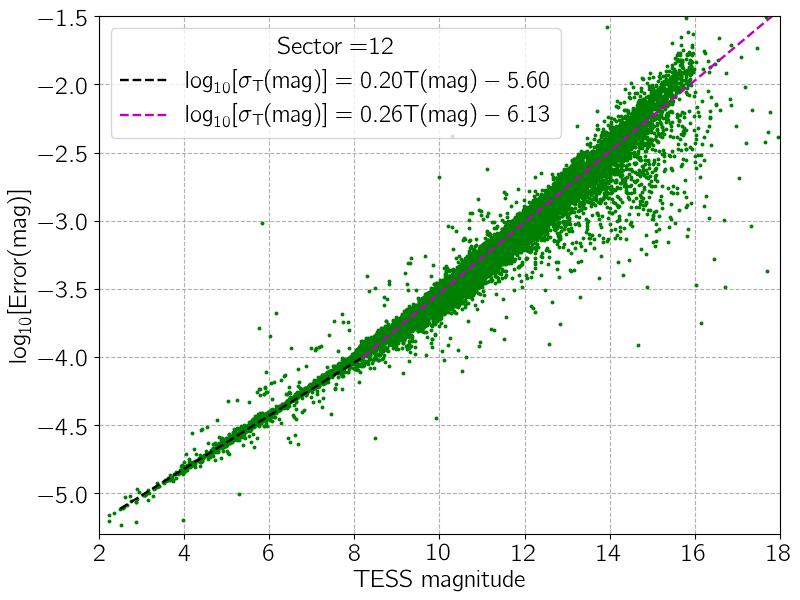}
\includegraphics[width=0.49\textwidth]{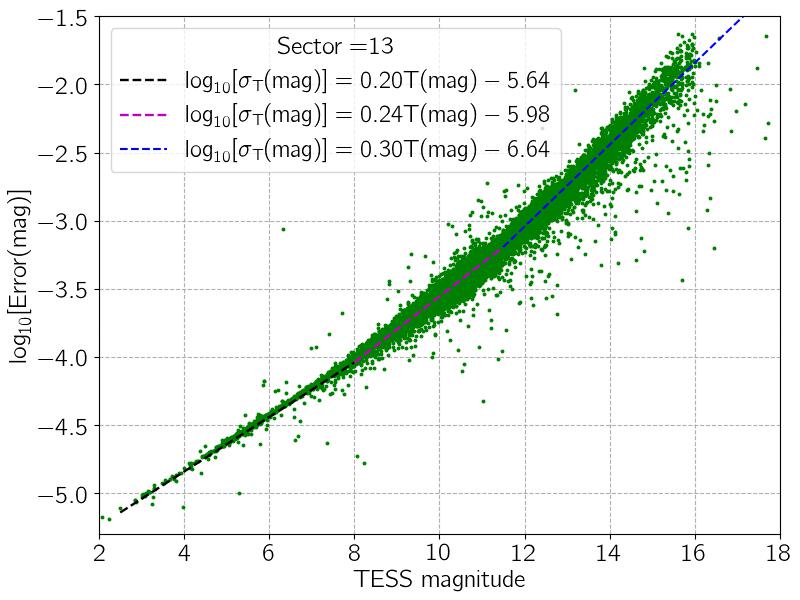}
\caption{Similar to Figure \ref{error}, but these source stars are inside FFIs taken from sectors $12$ and $13$. The stars parameters were extracted by the \texttt{TESS}-\texttt{SPOC} pipeline \citep{TESS-SCOPE}.}\label{error2}
\end{figure*}	
\begin{figure*}
\centering
\subfigure[]{\includegraphics[width=0.32\textwidth]{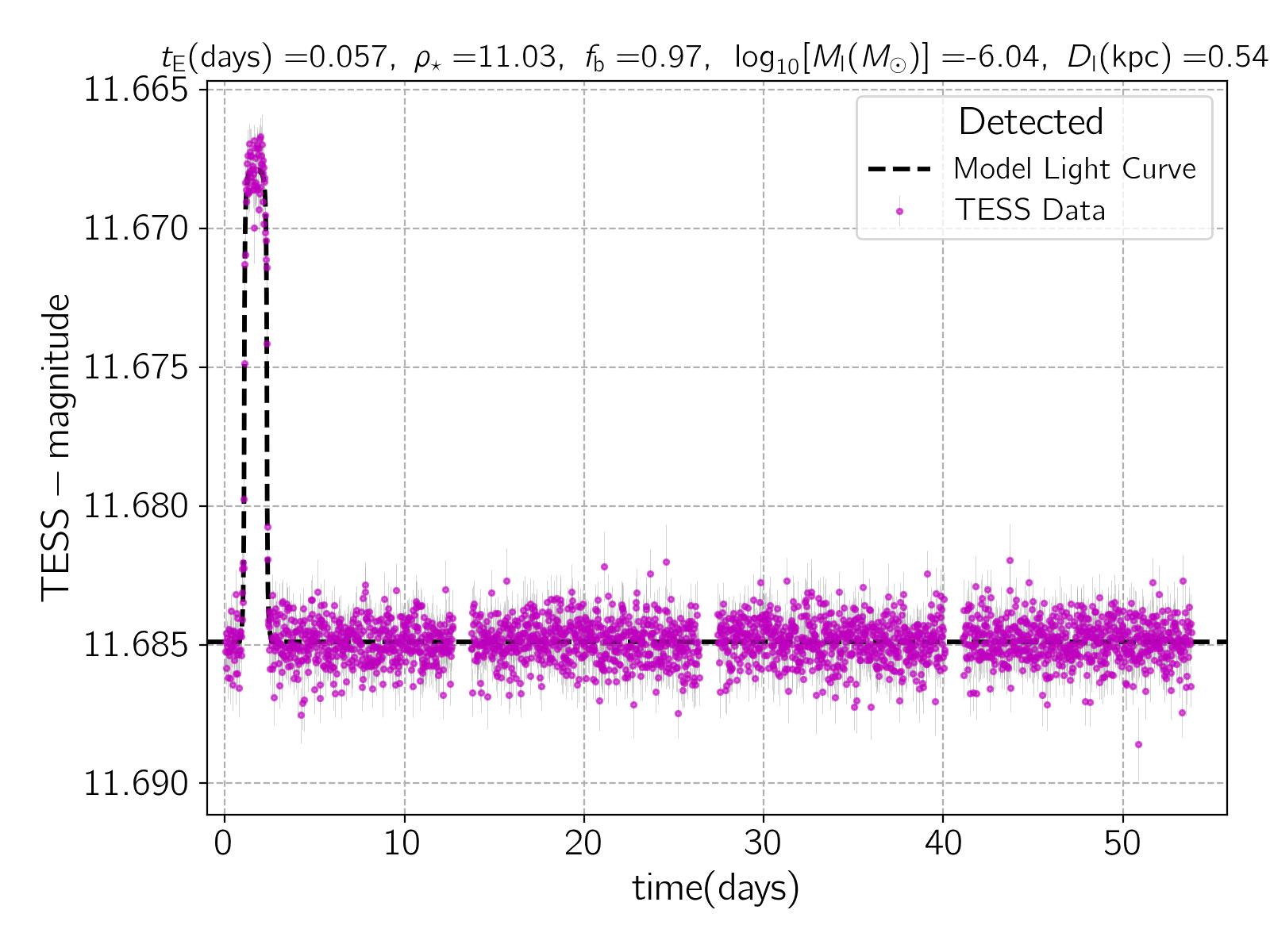}\label{light2a}}
\subfigure[]{\includegraphics[width=0.32\textwidth]{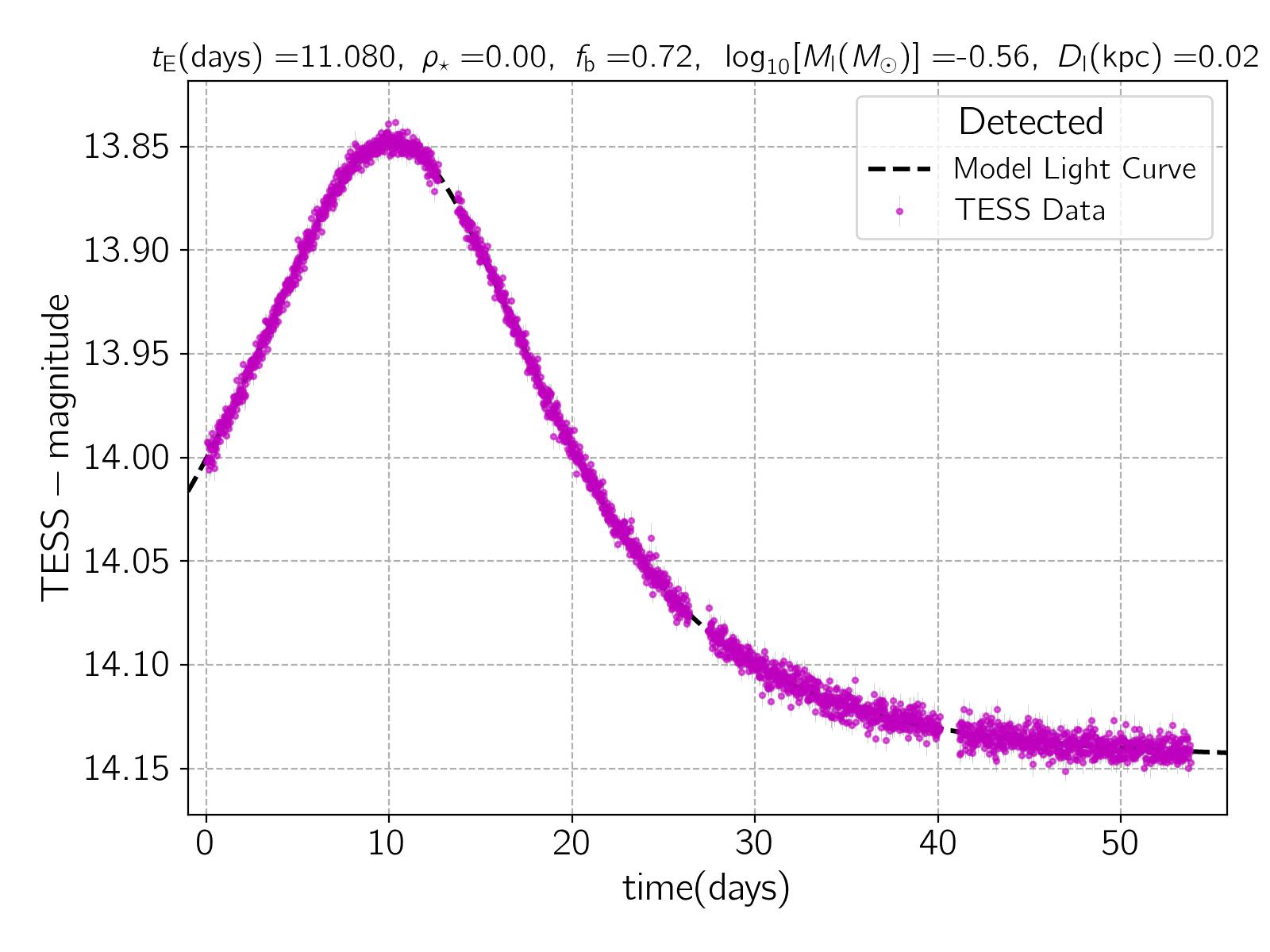}\label{light2b}}
\subfigure[]{\includegraphics[width=0.32\textwidth]{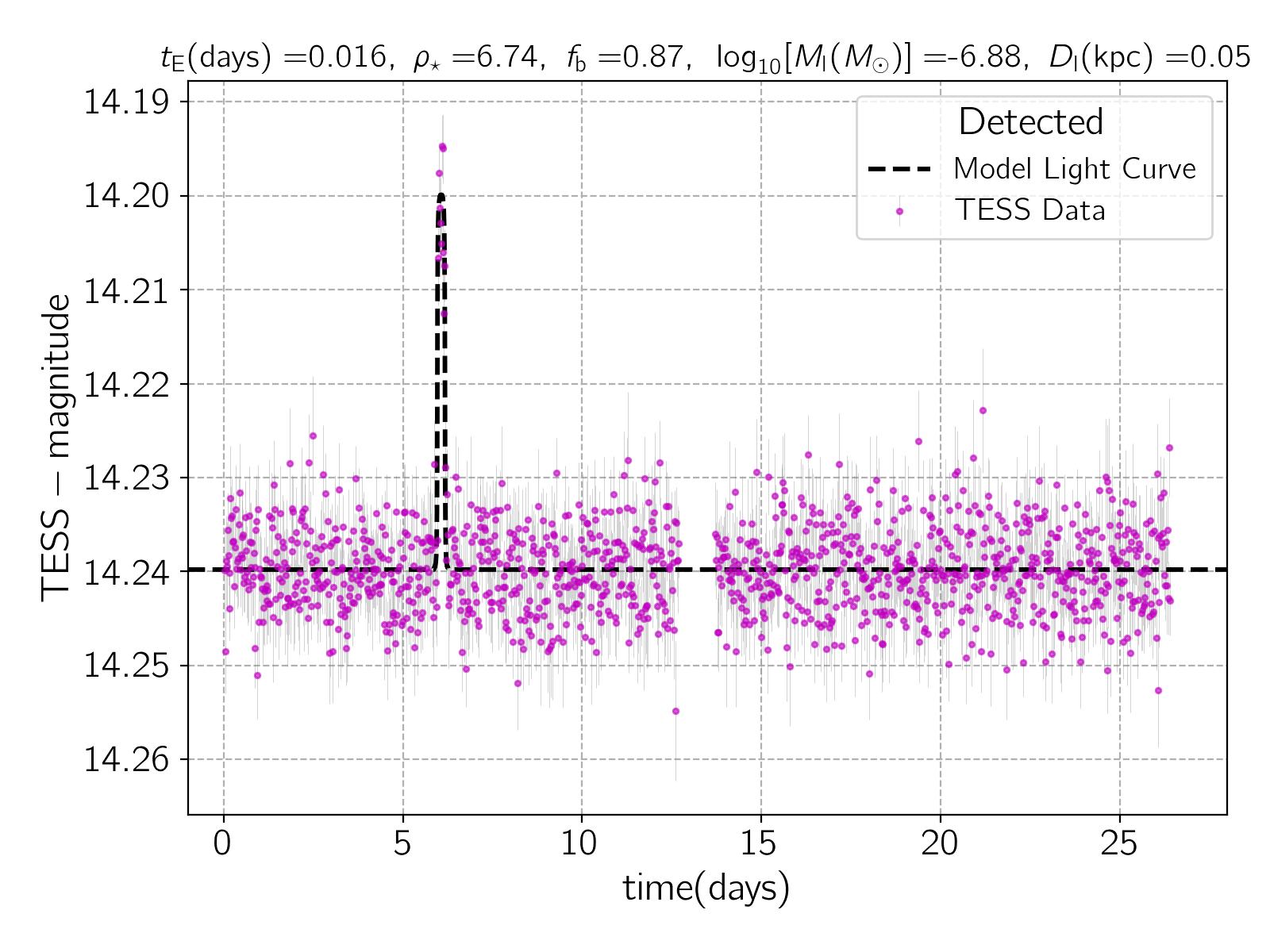}\label{light2c}}
\subfigure[]{\includegraphics[width=0.32\textwidth]{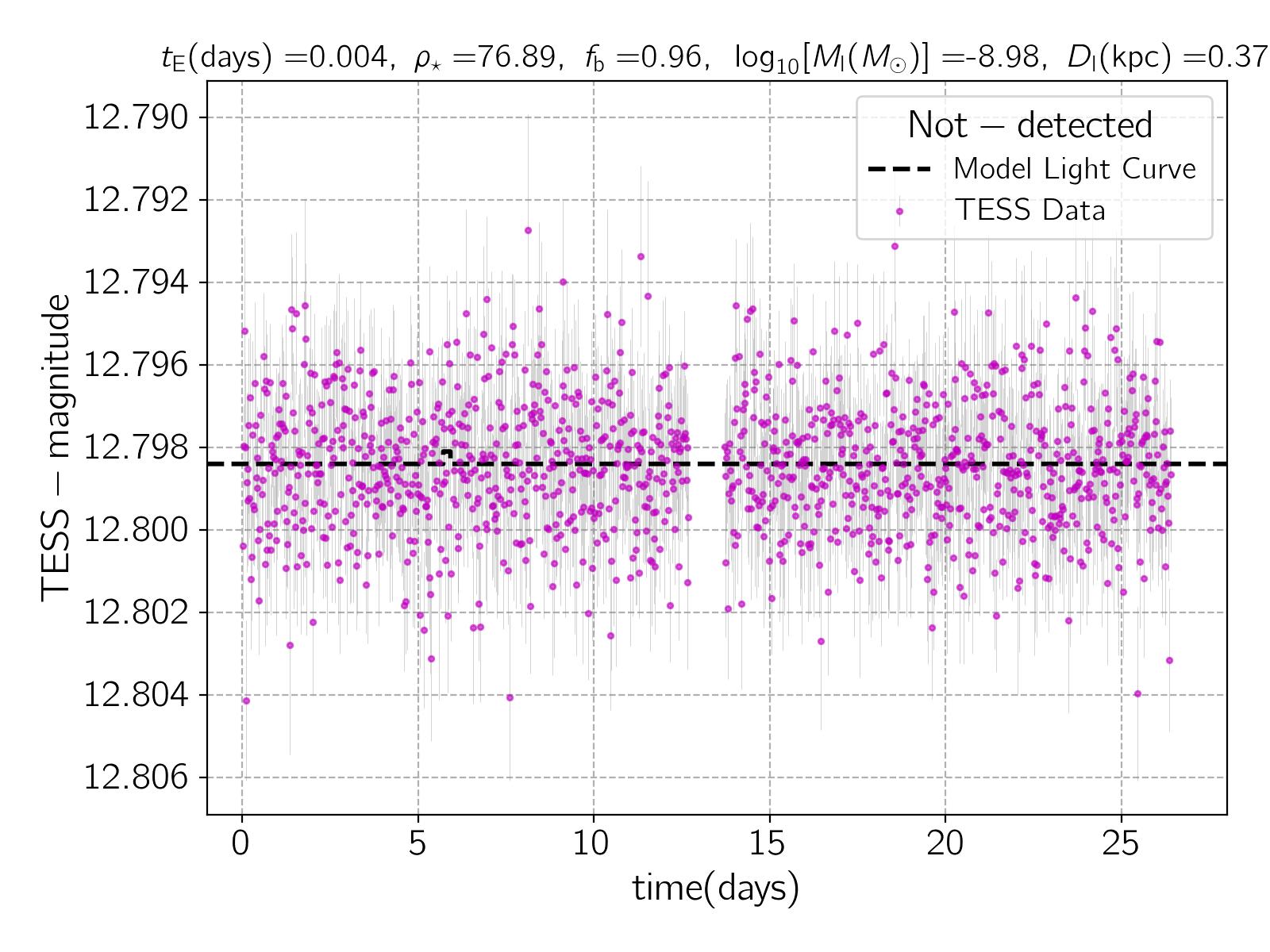}\label{light2d}}
\subfigure[]{\includegraphics[width=0.32\textwidth]{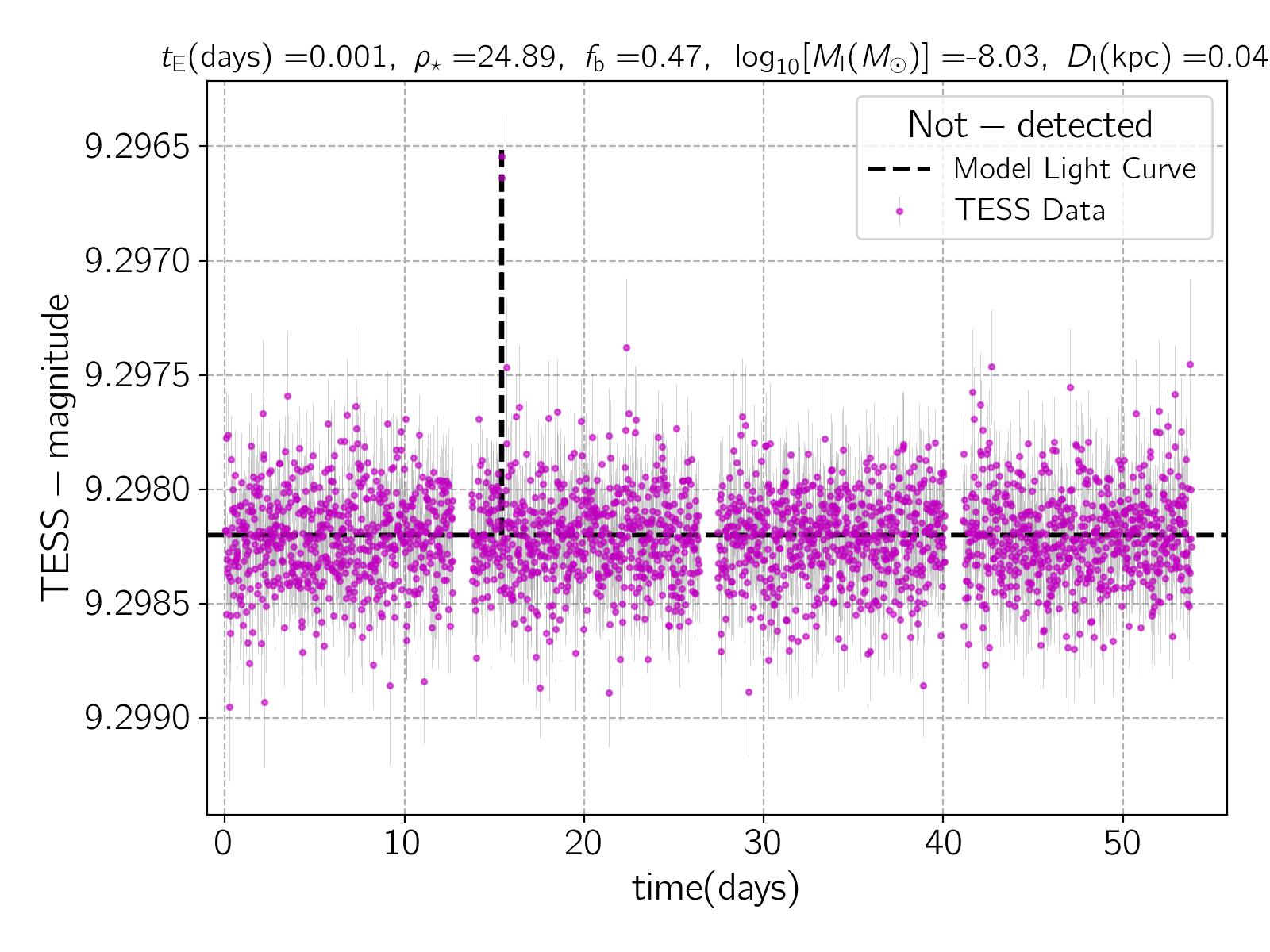}\label{light2e}}
\subfigure[]{\includegraphics[width=0.32\textwidth]{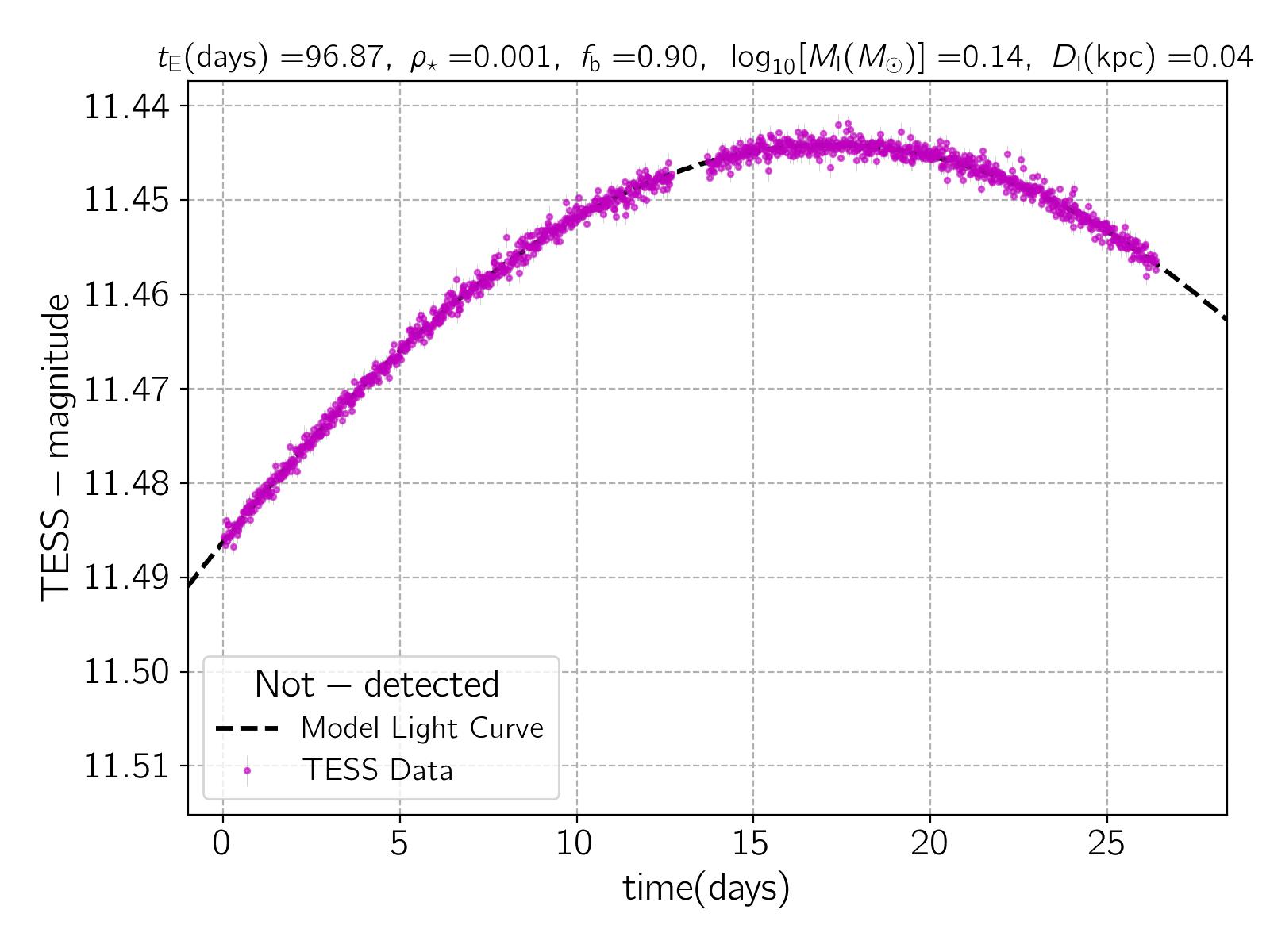}\label{light2f}}
\caption{Similar to Figure \ref{light}, but their source stars are the FFIs' targets. The observing cadence for simulated data is $30$ and $10$ minutes. Each gap between observing windows lasts one day, and each observing window lasts $12.7$ days.}\label{light2}
\end{figure*}

\section{Simulating microlensing events from the TESS FFIs' stars}\label{FFI}
In addition to the CTLs, the TESS community has released (and releases) the TESS Full-Frame Images (FFIs) which were taken every $30$, $10$, or $3.3$ minutes. The stellar light curves from FFIs resulting the TESS observations have been extracted using different pipelines \citep[e.g., ][]{2020HuangFFI}. In this work, we use the catalog of stars extracted from FFIs by the \texttt{TESS}-\texttt{SPOC} (TESS Science Processing Operations Center \citet{Jenkins2016}) pipeline which was well offered by \citet{TESS-SCOPE,tessspocstars}. Accordingly, for each sector and for $\lesssim160,000$ stars their light curves and some of their physical properties (including the stellar apparent magnitude, blending factor, effective temperature, surface gravity, metallicity, and stellar radius) were publicly published. The stellar distances were not determined in these mentioned ensembles.

To determine the stellar distance based on their indicated physical parameters, for each source star we find the most-similar star from the Besan\c{c}on model and take their absolute magnitudes for the FFIs' targets. In the Besan\c{c}on model, the stellar absolute magnitudes in the Gaia bands ($G$,~$G_{\rm BP}$,~$G_{\rm RP}$) are available. Additionally, \citet{2018AJstassun,2019AJstassun} have offered the relations between stellar absolute magnitudes in the Gaia bands and the TESS passband $T$. So, we determine the absolute magnitudes in the TESS passband of the FFIs' targets using those given relations. To determine their distance, we need their extinctions versus distance. For all lines of sight with the Galactic latitudes and longitudes $b\in [-10,~10]$ deg and $l\in [-100,~100]$ deg, we use the 3D extinction map (in the $K_{s}$ passband) offered by \citet{Marshal2006}. For other lines of sight, we use the extinctions in the standard $V$-band versus the distance from the observer using $A_{V}=0.7~D (\rm{kpc})$ which was proposed by the Besan\c{c}on model. We use the relation between the extinctions in different bands as given in \citet{Cardelli1989}. Accordingly, in the simulation we use $A_{T}= 0.62~A_{V}$, and $A_{V}=8.21~A_{K_{s}}$ to calculate the extinction in the TESS passband $T$. Finally, we estimate the distances of the TESS FFIs' stars based on their absolute and apparent magnitudes, and the extinction versus distance.

\begin{deluxetable*}{c c c c c c c c c c}
\tablecolumns{10}
\centering
\tablewidth{0.99\textwidth}\tabletypesize\footnotesize
\tablecaption{Similar to Table \ref{tab1} but for simulated microlensing events from the TESS FFIs' stars which are observed during a $27.4$-day observing time with a one-day gap in the middle.\label{tab3}}
\tablehead{\colhead{$\rm{sector}~\rm{No.}$}&\colhead{$D_{\rm l}$}&\colhead{$D_{\rm s}$}&\colhead{$\log_{10}[M_{\rm l}]$}&\colhead{$\log_{10}[t_{\rm E}]$}&\colhead{$\log_{10}[R_{\rm E}]$}&\colhead{$v_{\rm{rel}}$}&\colhead{$\rho_{\star}$}&\colhead{$\log_{10}[\pi_{\rm E}]$}&\colhead{$m_{\rm{TESS}}$}\\
& $\rm{kpc}$&$\rm{kpc}$&$M_{\odot}$& $\rm{days}$ & $\rm{au}$ & $\rm{km}/s$ & & &$\rm{mag}$}
\startdata
$1$ & $0.07,~0.23$ & $0.21,~0.52$ & $-6.39,~-5.87$ & $-1.74,~-1.21$ & $-3.38,~-3.06$ & $21.1,~51.1$ & $1.7,~5.5$ & $2.66,~3.14$ & $11.6,~12.9$\\
$2$ & $0.07,~0.23$ & $0.24,~0.55$ & $-6.39,~-5.89$ & $-1.73,~-1.20$ & $-3.37,~-3.05$ & $21.8,~51.3$ & $1.6,~5.1$ & $2.67,~3.13$ & $11.8,~13.0$\\
$3$ & $0.06,~0.22$ & $0.23,~0.52$ & $-6.38,~-5.85$ & $-1.73,~-1.19$ & $-3.37,~-3.05$ & $21.1,~51.6$ & $1.5,~4.9$ & $2.67,~3.13$ & $11.7,~13.1$\\
$4$ & $0.07,~0.24$ & $0.25,~0.54$ & $-6.38,~-5.87$ & $-1.68,~-1.15$ & $-3.36,~-3.04$ & $19.4,~46.8$ & $1.7,~5.3$ & $2.66,~3.13$ & $11.6,~12.9$\\
$5$ & $0.07,~0.26$ & $0.26,~0.58$ & $-6.38,~-5.88$ & $-1.61,~-1.08$ & $-3.35,~-3.04$ & $16.8,~40.5$ & $1.9,~6.0$ & $2.62,~3.10$ & $11.2,~12.5$\\
$6$ & $0.08,~0.31$ & $0.25,~0.63$ & $-6.40,~-5.89$ & $-1.57,~-1.03$ & $-3.35,~-3.01$ & $15.3,~36.3$ & $2.5,~8.2$ & $2.58,~3.08$ & $10.5,~11.7$\\
$7$ & $0.09,~0.30$ & $0.29,~0.61$ & $-6.39,~-5.89$ & $-1.59,~-1.05$ & $-3.32,~-3.02$ & $16.1,~40.7$ & $2.5,~7.7$ & $2.58,~3.07$ & $10.7,~11.9$\\
$8$ & $0.08,~0.27$ & $0.25,~0.58$ & $-6.38,~-5.87$ & $-1.67,~-1.14$ & $-3.34,~-3.04$ & $19.2,~46.6$ & $2.0,~6.4$ & $2.61,~3.08$ & $11.3,~12.5$\\
$9$ & $0.07,~0.25$ & $0.25,~0.54$ & $-6.39,~-5.86$ & $-1.72,~-1.19$ & $-3.35,~-3.04$ & $21.5,~50.4$ & $1.8,~5.7$ & $2.63,~3.10$ & $11.4,~12.7$\\
$10$ & $0.07,~0.25$ & $0.25,~0.54$ & $-6.40,~-5.88$ & $-1.71,~-1.17$ & $-3.36,~-3.05$ & $19.4,~48.1$ & $2.0,~6.2$ & $2.63,~3.11$ & $11.3,~12.6$\\
$11$ & $0.08,~0.27$ & $0.26,~0.54$ & $-6.39,~-5.89$ & $-1.65,~-1.11$ & $-3.36,~-3.05$ & $17.3,~42.8$ & $2.3,~7.5$ & $2.60,~3.08$ & $10.7,~12.2$\\
$12$ & $0.07,~0.33$ & $0.20,~0.64$ & $-6.39,~-5.86$ & $-1.58,~-1.03$ & $-3.37,~-3.00$ & $15.1,~36.5$ & $3.0,~10.1$ & $2.55,~3.09$ & $10.2,~11.5$\\
$13$ & $0.07,~0.26$ & $0.23,~0.54$ & $-6.39,~-5.88$ & $-1.62,~-1.09$ & $-3.38,~-3.06$ & $16.0,~37.2$ & $2.7,~8.7$ & $2.60,~3.10$ & $10.2,~11.9$\\
$14$ & $0.08,~0.29$ & $0.25,~0.58$ & $-6.39,~-5.87$ & $-1.68,~-1.11$ & $-3.36,~-3.01$ & $18.2,~45.8$ & $2.6,~8.7$ & $2.57,~3.06$ & $10.3,~11.7$\\
$15$ & $0.08,~0.31$ & $0.26,~0.60$ & $-6.38,~-5.87$ & $-1.68,~-1.11$ & $-3.34,~-3.02$ & $18.3,~46.7$ & $2.7,~8.3$ & $2.56,~3.06$ & $10.5,~11.8$\\
$16$ & $0.09,~0.31$ & $0.28,~0.62$ & $-6.38,~-5.87$ & $-1.65,~-1.10$ & $-3.32,~-3.00$ & $18.4,~47.3$ & $2.6,~8.2$ & $2.56,~3.05$ & $10.5,~11.8$\\
$17$ & $0.07,~0.25$ & $0.23,~0.53$ & $-6.38,~-5.86$ & $-1.70,~-1.15$ & $-3.37,~-3.04$ & $19.4,~46.9$ & $1.9,~6.0$ & $2.64,~3.12$ & $11.2,~12.5$\\
$18$ & $0.08,~0.29$ & $0.26,~0.58$ & $-6.38,~-5.88$ & $-1.61,~-1.08$ & $-3.34,~-3.03$ & $17.1,~40.8$ & $2.2,~7.2$ & $2.59,~3.08$ & $10.9,~12.2$\\
$19$ & $0.09,~0.35$ & $0.27,~0.75$ & $-6.38,~-5.88$ & $-1.53,~-0.98$ & $-3.32,~-2.98$ & $14.7,~36.0$ & $2.4,~7.4$ & $2.56,~3.06$ & $10.8,~12.0$\\
$20$ & $0.08,~0.27$ & $0.26,~0.58$ & $-6.38,~-5.87$ & $-1.59,~-1.07$ & $-3.34,~-3.02$ & $17.0,~39.0$ & $2.1,~6.5$ & $2.60,~3.09$ & $11.0,~12.2$\\
$21$ & $0.07,~0.24$ & $0.24,~0.55$ & $-6.38,~-5.88$ & $-1.68,~-1.16$ & $-3.36,~-3.05$ & $19.7,~46.6$ & $1.8,~5.5$ & $2.66,~3.12$ & $11.5,~12.8$\\
$22$ & $0.06,~0.22$ & $0.24,~0.55$ & $-6.38,~-5.88$ & $-1.73,~-1.21$ & $-3.37,~-3.05$ & $21.2,~49.7$ & $1.5,~4.8$ & $2.69,~3.16$ & $11.7,~13.1$\\
$23$ & $0.07,~0.23$ & $0.23,~0.55$ & $-6.39,~-5.87$ & $-1.71,~-1.20$ & $-3.36,~-3.04$ & $20.7,~51.7$ & $1.6,~4.9$ & $2.68,~3.12$ & $11.8,~13.1$\\
$24$ & $0.07,~0.24$ & $0.23,~0.52$ & $-6.39,~-5.86$ & $-1.70,~-1.19$ & $-3.36,~-3.05$ & $20.4,~47.9$ & $1.7,~5.5$ & $2.65,~3.11$ & $11.6,~13.0$\\
$25$ & $0.07,~0.24$ & $0.24,~0.52$ & $-6.38,~-5.87$ & $-1.66,~-1.15$ & $-3.37,~-3.05$ & $18.1,~45.0$ & $1.9,~6.0$ & $2.65,~3.13$ & $11.2,~12.5$\\
$26$ & $0.08,~0.28$ & $0.25,~0.53$ & $-6.40,~-5.87$ & $-1.66,~-1.11$ & $-3.36,~-3.04$ & $17.4,~44.6$ & $2.4,~7.5$ & $2.60,~3.08$ & $10.7,~12.1$\\
\enddata
\tablecomments{The first column represents the secrtor number.}
\end{deluxetable*}

The TESS photometric errors for these stars is somewhat different from those due to the TESS CTLs (as shown in Figure \ref{error}). For these stars and for each sector we first fit linear relations between the median photometric errors and the stellar apparent magnitudes of stars extracted from the \texttt{TESS-SPOC} pipeline. Two examples due to sectors 12 and 13 are shown in Figure \ref{error2}. The best-fitted linear lines to these data are plotted in these panels with dashed lines. For other sectors, the best-fitted lines can be found in the Zenodo address \citet{sajadian2024zenodo}. Other details of these Monte-Carlo simulations are the same as ones explained in the previous section. 

Some examples of simulated events are shown in Figure \ref{light2}. In these plots, the observing cadences are $30$ minutes for all light curves. The observing windows are $12.7$ days and their gaps last one day. Three top panels represent the detectable events and three bottom ones show microlensing events for which the TESS data are not sufficient to be realized. The microlensing light curve \ref{light2e} is too short to be captured through observations with a $30$-minute cadence. Also the light curve \ref{light2d} is affected by extremely large finite-source size and it is not detectable. However, two light curves \ref{light2a}, and \ref{light2c} have considerable finites-source sizes  and both are due to FFPs and detectable. We label the light curve \ref{light2f} as a not-detectable one, because there is no data covering the baseline.  

We have performed the Monte-Carlo simulation for sectors 1, 2, 3, ... 25, and 26 which were observed by the TESS during the first and second year of its primary mission. For all of these sectors we fix observing time to $27.4$ days with one $1$-day gap in the middle, since most of stars inside each sector ($74\%$) are observed during $27.4$ days. We mention the results of these simulations in Tables \ref{tab3} and \ref{tab4}. The format of these tables are similar to Tables \ref{tab1}, and \ref{tab2}, but each row represents the (lensing and statistical) information due to each sector, as the sector's number is mentioned in the first columns.

\begin{deluxetable*}{c c c c c c c c c c c c}
\tablecolumns{12}
\centering
\tablewidth{0.99\textwidth}\tabletypesize\footnotesize
\tablecaption{Similar to Table \ref{tab2}, but they are due to simulated microlensing events from the FFIs' targets. We assume these stars are observed during $27.4$ days with a $1$-day gap in the middle.\label{tab4}}
\tablehead{\colhead{$\rm{sector}~\rm{No.}$}&\colhead{$T_{\rm{obs}}$}&\colhead{$\overline{\tau} \times 10^{9}$}&\colhead{$\Gamma \times 10^{9}$}&\colhead{$\left<t_{\rm E}\right>$}&\colhead{$\mathcal{A}$}& \colhead{$\Gamma_{\rm{TESS}}\times 10^{9}$}&\colhead{$\hat{N_{\rm e}}\times 10^{8}$}&\colhead{$N_{\star}^{\dag}$}&\colhead{$N_{\rm e}$}&\colhead{$\epsilon$}& \colhead{$f_{\rm{MS}}:f_{\rm{BD}}:f_{\rm{FFP}}$}\\ 
&$\rm{days}$& &$\rm{star}^{-1}\rm{days}^{-1}$&$\rm{days}$& & $\rm{star}^{-1}\rm{days}^{-1}$& $\rm{star}^{-1}$ & & & $[\%]$ & $[\%]$}
\startdata
$1$ & $27.4$ & $0.80$ & $1.57$ & $0.32$ & $0.92$ & $1.44$ & $3.95$ & $118 k$ & $0.005$ & $91.8$ & $0.4:0.4:99.3$\\
$2$ & $27.4$ & $0.87$ & $1.68$ & $0.33$ & $0.91$ & $1.53$ & $4.18$ & $118 k$ & $0.005$ & $90.9$ & $0.4:0.4:99.3$\\
$3$ & $27.4$ & $0.76$ & $1.50$ & $0.32$ & $0.91$ & $1.36$ & $3.73$ & $118 k$ & $0.004$ & $90.9$ & $0.4:0.4:99.3$\\
$4$ & $27.4$ & $0.89$ & $1.60$ & $0.36$ & $0.92$ & $1.46$ & $4.01$ & $118 k$ & $0.005$ & $91.6$ & $0.3:0.3:99.4$\\
$5$ & $27.4$ & $1.21$ & $1.71$ & $0.45$ & $0.92$ & $1.58$ & $4.32$ & $118 k$ & $0.005$ & $92.1$ & $0.2:0.3:99.5$\\
$6$ & $27.4$ & $1.97$ & $2.50$ & $0.50$ & $0.93$ & $2.32$ & $6.35$ & $118 k$ & $0.007$ & $92.6$ & $0.2:0.3:99.5$\\
$7$ & $27.4$ & $1.73$ & $2.46$ & $0.45$ & $0.93$ & $2.29$ & $6.28$ & $118 k$ & $0.007$ & $93.1$ & $0.2:0.3:99.5$\\
$8$ & $27.4$ & $1.21$ & $2.01$ & $0.38$ & $0.92$ & $1.85$ & $5.08$ & $118 k$ & $0.006$ & $92.2$ & $0.3:0.3:99.4$\\
$9$ & $27.4$ & $1.05$ & $1.96$ & $0.34$ & $0.92$ & $1.80$ & $4.92$ & $118 k$ & $0.006$ & $91.6$ & $0.3:0.3:99.3$\\
$10$ & $27.4$ & $1.07$ & $2.05$ & $0.33$ & $0.91$ & $1.86$ & $5.10$ & $118 k$ & $0.006$ & $90.5$ & $0.3:0.3:99.4$\\
$11$ & $27.4$ & $1.40$ & $2.30$ & $0.39$ & $0.93$ & $2.13$ & $5.85$ & $118 k$ & $0.007$ & $92.6$ & $0.3:0.3:99.4$\\
$12$ & $27.4$ & $2.80$ & $3.93$ & $0.45$ & $0.90$ & $3.53$ & $9.68$ & $118 k$ & $0.011$ & $89.9$ & $0.3:0.3:99.4$\\
$13$ & $27.4$ & $1.55$ & $2.36$ & $0.42$ & $0.91$ & $2.15$ & $5.88$ & $118 k$ & $0.007$ & $91.0$ & $0.3:0.3:99.4$\\
$14$ & $27.4$ & $2.07$ & $3.38$ & $0.39$ & $0.90$ & $3.05$ & $8.36$ & $118 k$ & $0.010$ & $90.2$ & $0.3:0.3:99.4$\\
$15$ & $27.4$ & $1.95$ & $3.17$ & $0.39$ & $0.90$ & $2.86$ & $7.84$ & $118 k$ & $0.009$ & $90.1$ & $0.3:0.3:99.3$\\
$16$ & $27.4$ & $1.96$ & $3.15$ & $0.40$ & $0.90$ & $2.85$ & $7.80$ & $118 k$ & $0.009$ & $90.5$ & $0.3:0.3:99.4$\\
$17$ & $27.4$ & $1.04$ & $1.80$ & $0.37$ & $0.92$ & $1.64$ & $4.51$ & $118 k$ & $0.005$ & $91.6$ & $0.3:0.3:99.4$\\
$18$ & $27.4$ & $1.39$ & $2.04$ & $0.43$ & $0.92$ & $1.87$ & $5.13$ & $118 k$ & $0.006$ & $91.6$ & $0.3:0.3:99.4$\\
$19$ & $27.4$ & $2.44$ & $2.95$ & $0.53$ & $0.93$ & $2.74$ & $7.49$ & $118 k$ & $0.009$ & $92.6$ & $0.2:0.3:99.5$\\
$20$ & $27.4$ & $1.34$ & $1.87$ & $0.46$ & $0.92$ & $1.71$ & $4.68$ & $118 k$ & $0.006$ & $91.5$ & $0.2:0.3:99.5$\\
$21$ & $27.4$ & $0.93$ & $1.63$ & $0.36$ & $0.91$ & $1.48$ & $4.07$ & $118 k$ & $0.005$ & $90.9$ & $0.3:0.3:99.4$\\
$22$ & $27.4$ & $0.79$ & $1.52$ & $0.33$ & $0.90$ & $1.38$ & $3.77$ & $118 k$ & $0.004$ & $90.3$ & $0.4:0.4:99.3$\\
$23$ & $27.4$ & $0.85$ & $1.65$ & $0.33$ & $0.90$ & $1.48$ & $4.05$ & $118 k$ & $0.005$ & $89.6$ & $0.4:0.4:99.3$\\
$24$ & $27.4$ & $0.91$ & $1.70$ & $0.34$ & $0.89$ & $1.51$ & $4.15$ & $118 k$ & $0.005$ & $89.0$ & $0.3:0.3:99.3$\\
$25$ & $27.4$ & $1.06$ & $1.89$ & $0.36$ & $0.91$ & $1.72$ & $4.70$ & $118 k$ & $0.006$ & $90.8$ & $0.3:0.3:99.4$\\
$26$ & $27.4$ & $1.43$ & $2.35$ & $0.39$ & $0.91$ & $2.15$ & $5.88$ & $118 k$ & $0.007$ & $91.1$ & $0.3:0.3:99.4$\\
\enddata
\tablecomments{$^{\dag}$We adjusted the details of the Monte Carlo simulations with the FFIs' stars extracted by the \citet{TESS-SCOPE} pipeline. In this reference, for each sector the light curves and physical parameters for $\lesssim 160,000$ stars were released. On average, $74\%$ of these stars (around $118,000$) were observed by the TESS telescope during $27.4$ days.}
\end{deluxetable*}

According to Tables \ref{tab3} and \ref{tab4}, for the TESS FFIs' stars the microlensing optical depth is on average $5$-$15$ times larger than the optical depth due to the TESS CTL stars, because the FFIs' stars are on average farther ($2.5$-$3$ times). The microlensing event rate for the FFIs' stars is on average $2$-$6$ times larger than the event rate due to the CTL stars. The potential microlensing events from the FFIs' stars are on average longer and are on average due to more massive lens objects. The fractions of FFPs in detectable events from the FFI and CTL are $99\%$, and the probability of detecting FFPs from analyzing stellar light curves extracted from the TESS FFIs is somewhat higher than that from the CTL light curves. The microlensing events from the FFIs stars are somewhat less affected by the parallax effect.

The efficiencies for detecting microlensing events from the FFIs and CTL are $90\%$, and $95\%$, respectively. We note that the FFI stars are on average fainter than the CTL stars by $\sim 1.5$ mag in the TESS passband. Since the number of events depends on the number of background stars that the TESS will detect by the end of its mission and the applied pipeline to extract the source stars, so we compare the number of events per star (during $27.4$-day observations) for FFIs and CTL targets which are $\hat{N_{\rm e}}(\rm{star}^{-1})\simeq 4-10\times 10^{-8}$, and $1.6 \times 10^{-8}$, respectively. Hence, per star, we expect on average higher number of microlensing events from the FFIs stars than those from the TESS CTL stars.  

When comparing results from different sectors that have the same observing durations, we find that the maximum and minimum microlensing optical depths are due to sectors $12$ and $3$, respectively. We note that sector $12$ covers some part of the Galactic disk which is close to the Galactic bulge where the crowdedness of stars is high. The highest event rate is also due to the sector $12$ which is $\sim 6$ times higher than the event rate due to the CTL targets.

The number of microlensing events per star, $\hat{N}_{\rm e}$, is reported in the eighth column of Table \ref{tab4}. Accordingly, for all sectors 1 to 26 and during different $27.4$-day observing time spans, the overall number of microlensing events per star is $\hat{N}_{\rm e, \rm{tot}}\sim 1.4\times 10^{-6}$. Hence, to detect at least one microlensing event for the FFIs' stars $\sim720,000$ stellar light curves should be searched. The potential microlensing events for the TESS FFIs' stars are toward the Galactic disk, and $\sim 99\%$ of them are due to nearby FFIs.

We note that \citet{2020HuangFFI} extracted $\sim 25$ million stellar (magnitude-limited) light curves from the TESS FFIs released after the first two years of its mission. Nevertheless, we did not repeat Monte-Carlo simulations for the FFIs' stars extracted by \citet{2020HuangFFI} to determine a more accurate value for the expected number of microlensing events, because the blending factor was not reported for the stars in this catalog. Considering the large size of stellar PSF in the TESS observations, the blending effect for each source star is considerable and we could not ignore this effect in the simulations.

In the tenth column of Table \ref{tab4}, we report the expected number of microlensing events, $N_{\rm e}$, which are based on the stellar catalog extracted from the TESS FFIs using the \texttt{TESS-SPOC} pipeline. In this catalog, for each sector only up to $160,000$ stellar light curves were extracted, where $118,000$ of these stars were detected during $27.4$ days. By the end of the TESS mission the number of stellar light curves extracted from the TESS FFIs will be even much more which results higher number of microlensing events.

\section{Conclusions}\label{conclu}
The TESS telescope, designed by the NASA Explorer mission, was planned to detect the transiting planets which are rotating the near and bright stars. Meanwhile, other astrophysical and transient/periodic events would be captured through the TESS observations. This could include gravitational microlensing. In this work, we statistically studied detection of gravitational microlensing events in the TESS public data by calculating the optical depth, the microlensing event rate, and the number of microlensing events per observing star. Additionally, by performing Monte-Carlo simulations from all possible microlensing events and extracting detectable ones we investigated the physical and lensing properties of detectable events. There are two kinds of source stars for potential microlensing events detectable by the TESS telescope, which are the TESS CTL stars and the TESS FFIs' stars.

We took a sample of the TESS CTLs and simulated their potential microlensing events due to all possible lens objects (from main-sequences to low-mass FFPs in the wide mass range $M_{\rm l}\in [0.1M_{\oplus},~2M_{\odot}]$). In the simulations, we took the mass of lens objects from the mass function offered by \citet{Sumi2023} which was rewritten in Equation \ref{massf}. We extracted the discernible events based on four criteria which were (i) $\Delta \chi^{2}>500$ where $\Delta \chi^{2}$ is the difference between $\chi^{2}$ values from fitting the real model and the baseline, (ii) at least three data points should be above the baseline by $5 \sigma_{\rm m}$ ($\sigma_{\rm m}$ is the photometric error),  (iii) at least five data points should cover the baseline (where the magnification factor is less than $1.1$), and (iv) each side of microlensing light curves should be covered by at least three data points. 

We concluded that the microlensing optical depth for the CTL stars is on average $\simeq 0.2 \times 10^{-9}$ which is three orders of magnitude less than the microlensing optical depth toward the Galactic bulge. The TESS CTLs are mostly close to the observer, and their distances are less than $\sim 0.2$ kpc which results small microlensing optical depth. The microlensing event rate for these targets is on average $\Gamma_{\rm{TESS}}\simeq 0.6\times 10^{-9}$. The optical depth is a decreasing function versus the CTL priority, since the stellar distances from the observer decrease versus the CTL priority. Nevertheless, the efficiency for detecting microlensing events enhances with the CTL priority. We found that the highest efficiency for detecting the TESS microlensing events ($\sim 100\%$) happens for the lens mass in the range $10^{-4.5}-10^{-2.4} M_{\odot}$ (i.e., super-Earth to up Jupiter-mass FFPs), and on average $\sim 99\%$ of these detectable microlensing events are due to FFPs. We concluded that the total expected number of microlensing events from the TESS CTLs $N_{\rm{e}, \rm{tot}}\sim 0.03$, which means the chance for detecting microlensing events from the TESS observations of the CTL targets (based on our detectability criteria) is too low.

Additionally, we simulated potential microlensing events from the FFIs' stars by fixing the observing time to $27.4$ days. We adjusted our simulations for the FFIs' stars extracted from \texttt{TESS-SPOC} pipeline \citep{TESS-SCOPE} and performed the simulation for each sector (from 1 to 26) separately. For the stars in this catalog the necessary parameters to simulate their microlensing events, e.g., the blending factor, effective surface temperature, surface gravity, radius, etc., were reported. Generally the stars from the TESS FFIs are on average fainter than CTLs (by $\sim 1.5$ mag in the TESS passband). Hence, their microlensing events have lower detection efficiencies, i.e., $\sim 90\%$, in comparison to the detection efficiency for events from the TESS CTL stars which is $\sim 94\%$. Detectable microlensing events from the TESS FFIs' stars have the average time scale (relative velocity-weighted) $\left<t_{\rm E}\right>\simeq 0.3$-$0.5$ day, and $99\%$ of them are due to FFPs. 

The microlensing optical depth for these targets is higher $\simeq 1$-$3\times 10^{-9}$. Their event rate is on average $\Gamma_{\rm{TESS}}\simeq 1$-$4 \times 10^{-9}$. The maximum (and minimum) optical depth and event rate happen for stars inside the sector $12$ which covers some part of the Galactic disk close to the Galactic bulge (and sector 3). The expected number of microlensing events per star from FFIs stars (inside sectors 1 to 26 and during different $27.4$-day observing windows) is $\hat{N}_{\rm{e}, \rm{tot}}\simeq 1.4 \times 10^{-6}$. Hence, to find at least one microlensing event among different $27.4$-day observations $\sim700,000$ stellar light curves extracted from FFIs should be searched. Such microlensing events are due to FFPs with a high probability. We note that \citet{2020HuangFFI} extracted the light curves of $\sim 25$ million stars from the TESS FFIs released after two first years of its mission, and by the end of the TESS mission the number of extracted stellar light curves from the TESS FFIs would be even higher.\\

\small
All simulations that have been done for this paper are available at: \url{https://github.com/SSajadian54/TESS_MicrolensingS}.  Additionally, they have been deposited to Zenodo \citep{sajadian2024zenodo}.\\

\small{
In this paper and in its Monte-Carlo simulations, we use the TESS CTL (with DOI number:  \url{doi:10.17909/fwdt-2x66}) and FFIs stars extracted by the \texttt{TESS}-\texttt{SPOC} pipeline (with DOI number: \url{doi:10.17909/t9-wpz1-8s54}) which were collected by the TESS mission that are publicly available from the MAST. Funding for the TESS mission is provided by NASA's Science Mission directorate. We acknowledge the use of TESS Alert data, which is currently in a beta test phase, from pipelines at the TESS Science Office and at the TESS Science Processing Operations Center. 
The authors gratefully thank T.~Barclay for kindly sharing his simulated data on numbers of detectable stars from the TESS FFIs. We thank K.~Stassun for his useful comments about the TESS photometric uncertainties. We also thank the anonymous referee for his/her careful and useful comments, which improved the quality of the paper.}\\

\bibliographystyle{aasjournal}
\bibliography{paper}{}
\end{document}